\documentclass[12pt,preprint]{aastex}        

\shorttitle{Br_$gamma$}
\shortauthors{Y. Li}

\begin{document}
\bibliographystyle{plainnat}

\title{Star Formation Rates in Resolved Galaxies: Calibrations with Near and Far Infrared Data for NGC\,5055 and NGC\,6946
\footnote{Based on observations obtained with WIRCam, a joint project of CFHT, Taiwan, Korea, Canada, France, and the Canada-France-Hawaii Telescope (CFHT) which is operated by the National Research Council (NRC) of Canada, the Institute National des Sciences de l'Univers of the Centre National de la Recherche Scientifique of France, and the University of Hawaii.}}
\author{Accepted to ApJ on Apr. 4. 2013}
\author{Yiming Li\footnote{Dept. of Astronomy, University of Massachusetts, Amherst, MA 01003, yimingl@astro.umass.edu} , 
Alison F. Crocker\footnotemark[2] , Daniela Calzetti\footnotemark[2] , Christine D. Wilson\footnote{Department of Physics \& Astronomy, McMaster University, Hamilton, Ontario L8S 4M1, Canada} , 
Robert C. Kennicutt\footnote{Institute of Astronomy, University of Cambridge, Madingley Road, Cambridge CB3 0HA, UK} , 
Eric J. Murphy\footnote{Observatories of the Carnegie Institution for Science, 813 Santa Barbara Street, Pasadena, CA 91101, USA} ,  Bernhard R. Brandl\footnote{Leiden Observatory, Leiden University, P.O. Box 9513, 2300 RA Leiden, The Netherlands} , B. T. Draine\footnote{Department of Astrophysical Sciences, Princeton University, Princeton, NJ 08544, USA} , M. Galametz\footnotemark[4] , B. D. Johnson\footnote{Institut d$^{\prime}$Astrophysique de Paris, UMR7095 CNRS, Universit{\'e} Pierre \& Marie Curie, 98 bis Boulevard Arago, 75014 Paris, France} , L. Armus\footnote{Spitzer Science Center, California Institute of Technology, MC 314-6, Pasadena, CA 91125, USA} , K. D. Gordon\footnote{Space Telescope Science Institute, 3700 San Martin Drive, Baltimore, MD 21218}\ \ $^{,}$\footnote{Sterrenkundig Observatorium, Universiteit Gent, Krijgslaan 281,B-9000 Gent, Belgium} ,  K. Croxall\footnote{Department of Physics and Astronomy, University of Toledo, Toledo, OH 43606, USA}\ \ $^{,}$\footnote{Department of Astronomy, The Ohio State University, 4051 McPherson Laboratory, Columbus, OH, 43210} , D. A. Dale\footnote{Department of Physics \& Astronomy, University of Wyoming, Laramie, WY 82071, USA} , C. W. Engelbracht\footnote{Steward Observatory, University of Arizona, Tucson, AZ 85721, USA}\ \ $^{,}$\footnote{Raytheon Company, 1151 E. Hermans Road, Tucson, AZ 85756, USA} , B. Groves\footnotemark[6]\ $^{,}$\footnote{Max-Planck-Institut f\"ur Astronomie, K\"onigstuhl 17, D-69117 Heidelberg, Germany} , C.-N. Hao\footnote{Tianjin Astrophysics Center, Tianjin Normal University, Tianjin 300387, China} , G. Helou\footnote{NASA Herschel Science Center, IPAC, California Institute of Technology, Pasadena, CA 91125, USA} , J. Hinz\footnotemark[15] , L. K. Hunt\footnote{INAF - Osservatorio Astroﬁsico di Arcetri, Largo E. Fermi 5, 50125 Firenze, Italy} , O. Krause\footnotemark[17] , H. Roussel\footnotemark[8] , M. Sauvage\footnote{CEA/DSM/DAPNIA/Service d$^{\prime}$Astrophysique, UMR AIM, CE Saclay, 91191 Gif sur Yvette Cedex} , J.D. T. Smith\footnotemark[12] }



\begin{abstract}
We use the near--infrared Br$\gamma$ hydrogen recombination line as a reference
 star formation rate (SFR) indicator to test the validity and establish the calibration of the
 {\it Herschel} PACS 70~$\mu$m emission as a SFR tracer
for sub--galactic regions in external galaxies. Br$\gamma$ offers the double advantage
of directly tracing ionizing photons and of being relatively insensitive
to the effects of dust attenuation. For our first experiment,
we use archival CFHT Br$\gamma$ and Ks images of two 
nearby galaxies: NGC\,5055 and NGC\,6946, which are also part of the  {\it Herschel}
program KINGFISH (Key Insights on Nearby Galaxies: a Far-Infrared Survey with Herschel). 
We use the extinction corrected Br$\gamma$ emission to derive the SFR(70) calibration for H{\sc ii} regions in these two galaxies.
A comparison of the SFR(70) calibrations at different spatial scales, from 200~pc to the size of the whole galaxy, reveals 
that about 50\% of the total 70$\mu$m emission is due to dust heated by stellar populations that are 
unrelated to the current star formation.
We use a simple model to qualitatively relate the increase of the SFR(70) calibration coefficient with
decreasing region size to the star formation timescale. We provide a calibration for an unbiased SFR indicator 
that combines the observed H$\alpha$ with the 70~$\mu$m emission, also for use in H{\sc ii} regions.
We briefly analyze the PACS 100 and 160~$\mu$m maps and find
that longer wavelengths are not as good SFR indicators as 70$\mu$m, in agreement with previous results.
We find that the calibrations show about 50\% difference between the two galaxies,
possibly due to effects of inclination.
\end{abstract}


\section{Introduction}
\label{intro}

Star formation rates (SFRs) have been measured and characterized 
across all wavelengths, from the X--ray, through the ultraviolet (UV), optical, infrared (IR), and all the way 
to the radio (for a review, see \citealt{kennicutt2012} for the past
decade, and \citealt{kennicutt1998} for earlier literature). 
Thanks to the synergy between   {\it Spitzer}, {\it GALEX} and {\it Herschel}, many SFR indicators have 
been recently defined and/or re--calibrated at UV, optical, and infrared wavelengths 
\citep[e.g.][]{wu2005, alonso2006, calzetti2007, calzetti2010, rieke2009, kennicutt2007, kennicutt2009, boquien2010, lawton2010, li2010, hao2011, murphy2011}.
However, any new calibration or re--calibration needs a {\em `reference'} SFR 
indicator against which it can be compared. Different authors use diverse {\em reference} SFR indicators (extinction--corrected H$\alpha$, 
total infrared, radio, combination of H$\alpha$ and 24~$\mu$m emission, etc.);
their choices are usually driven by the available datasets. Thus, newly (re)calibrated SFR tracers are not usually 
comparable, to each other or to more established indicators.
Furthermore, their accuracy and range of applicability often cannot be reliably established. For example, 
to calibrate an IR SFR indicator with a reference SFR indicator involving another IR band (e.g. the combination of H$\alpha$
and 24~$\mu$m), degeneracy may arise at high SFR regime where the dust emission dominates. 
A lack of a robust, wide field, reference SFR renders the absolute calibration and universality of the
24~$\mu$m term in hybrid tracers like H$\alpha+$24~$\mu$m and
FUV+24~$\mu$m uncertain, even for the well-studied
SINGS sample \citep{leroy2012}.
Progress in calibrating SFR 
indicators at a variety of wavelengths requires the use of `unbiased' reference SFR tracers, i.e. those insensitive to 
dust extinction, star formation history, etc. 

Near--infrared (NIR) hydrogen recombination lines, such as 
P$\alpha$ (1.8756~$\mu$m, from  {\it HST, Hubble Space Telescope}) and Br$\gamma$ (2.166~$\mu$m), and the radio free--free
emission \citep[e.g., with GBT, the Green Bank Telescope;][]{murphy2011, murphy2012}, trace the ionizing photons produced 
by massive stars and therefore the most recent ($<$10~Myr) star formation in a galaxy. 
As such, these tracers are less affected by variations in the star formation history of a region or galaxy
or, as in the case of infrared SFR tracers, by variations in the dust content \citep{kennicutt2012,calzetti2012b}. 
They are also significantly less affected by dust extinction than UV and H$\alpha$ emission; e.g. 
dust extinction is significantly lower at Br$\gamma$: A$_{H\alpha}$=1~mag 
corresponds to A$_{Br\gamma}$=0.14~mag. The NIR hydrogen recombination lines
are also free of in--band contamination from adjacent emission lines (e.g., [NII] for H$\alpha$ or non-thermal emission for the radio free-free emission).
P$\alpha$ and radio free--free emission have both previously been used as a reference SFR
to calibrate or study other SFR indicators \citep{calzetti2007, murphy2011}.
However, the use of these two indicators is limited due to the small field of view (FOV) of the P$\alpha$
observations with  {\it HST} (e.g. \citealt{calzetti2007} only studied the central 1--2~kpc of star--forming galaxies),
and the time-consuming nature of free--free observation 
to map whole galaxies.

The WIRCam (Wide Infrared Camera) on CFHT (Canada France Hawaii Telescope)
provides the capability to observe the NIR hydrogen recombination line, Brackett--$\gamma$ 
\citep[Br$\gamma$, 2.166~$\mu$m,][]{jones2002}, from the ground with large enough FOV 
(20$^{\prime}\times$20$^{\prime}$) to cover most nearby galaxies out to R$_{25}$ with one pointing (R$_{25}$ is defined as the 25 mag arcsec$^{-2}$ isophote). 
Much like P$\alpha$, Br$\gamma$ emission traces the ionizing photons from massive stars; 
it has the advantage of mapping efficiency over  {\it HST} P$\alpha$ or P$\beta$ (Wide Field Camera 3, WFC3) observations
 thanks to the large FOV of modern IR detectors. The lower angular resolution provided by ground-based observations relative to the  {\it HST} is, however, more than sufficient for our planned calibration of far--infrared emission.

High--sensitivity, high--angular resolution infrared telescopes, like the
 {\it Spitzer} and, more recently, the  {\it Herschel Space Telescopes},
have provided a major incentive to investigate dust--obscured
star formation in galaxies at all cosmic times, and at all scales, from 
galaxy--integrated to sub--galactic scales. Both telescopes have shown 
that dust--obscured starbursts dominate the star formation in the redshift 
range 1--3 \citep{lefloch2005, magnelli2009, elbaz2011, murphy2011b, reddy2012}. 
Concurrently, studies of nearby galaxies at a few hundred parsec (pc) resolution with 
 {\it Spitzer} and  {\it GALEX} have yielded new insights on the fundamental 
processes responsible for converting gas into stars and the Schmidt Law.  
Data finally probe the physical scales of Giant Molecular Associations 
\citep[$\sim$100-200 pc,][]{koda2009}, where models of star formation can be discriminated 
\citep{kennicutt2007, leroy2008, liu2011, rahman2012, calzetti2012}. Calibrations of SFR indicators in
 the far--infrared (FIR, $\gtrsim$40--50~$\mu$m) regime at sub-galactic scales have thus become useful for these kinds of studies, leading recent studies to focus on them \citep{calzetti2010, lawton2010, li2010}. 
The integrated FIR emission from 
 whole galaxies contains significant diffuse emission that comes from dust heated by stellar populations not related to the current star formation. A comparison in \citet{li2010} between the calibrations of SFR(70) for both sub-galactic regions and entire galaxies reveals that, on kiloparsec scales and larger, about 40\% of emission of 70~$\mu$m in the galaxies comes from diffuse emission, on average. 

In this paper, we use Br$\gamma$ maps of two KINGFISH \citep[Key Insights on Nearby Galaxies: a Far-Infrared Survey with Herschel,][]{kennicutt2011}
galaxies, retrieved from the CFHT archive: NGC\,5055 and NGC\,6946, 
 to calibrate the {\it Herschel}/PACS 70~$\mu$m band 
as a SFR indicator at sub-galactic scales. We also consider the 100 and 160~$\mu$m {\it Herschel} bands as SFR indicators.
The resolution of {\it Herschel} at 70~$\mu$m (FWHM$\sim$5.5$^{\prime\prime}$) enables us to resolve regions as small as 200~pc in these two galaxies. This scale is considerably smaller than a galaxy, 
and although several H{\sc ii} regions will be included, we expect the physical characteristics of the stellar population in these apertures to be different from those of the entire galaxy. This paper is devoted to quantifying the differences of the calibrations between the whole galaxies and sub--galactic regions, and to providing interpretation with model comparisons. 

Units in this paper are as follows: $\rm{ergs\cdot s^{-1}}$ for luminosity, $\rm{ergs\cdot s^{-1}\cdot kpc^{-2}}$ for luminosity surface density (LSD, with symbol $\rm\Sigma$), $\rm{M_{\odot}\cdot yr^{-1}}$ for SFR and $\rm{M_{\odot}\cdot yr^{-1}\cdot kpc^{-2}}$ for star formation rate surface density ($\rm\Sigma(SFR)$ or SFRD), unless otherwise specified. The infrared luminosity used in the paper of the IR bands is, L(IR) = $\nu L_{\nu}$, following the common definition of monochromatic IR flux.

\section{Data}
\label{data}

Table \ref{tab:data} contains a summary of the data involved in the analyses in this paper, along with their angular resolutions. Detailed descriptions are given in following sections.

\subsection{Archive CFHT Data}

Two galaxies, NGC\,5055 and NGC\,6946 (Table \ref{tab:gal}), out of the KINGFISH sample of 61 galaxies, have archival observations of
sufficient depth ($>$1000 sec integration time in the Br$\gamma$ narrowband, see Section \ref{calib} for the final noise levels of the reduced images) for our purposes with WIRCam \citep{puget2004}, observed by PI Daniel Devost with RUNID 07AD86 for NGC\,5055 and 07BD91 for NGC\,6946. WIRCam is the near infrared wide-field imager on the CFHT and has been in operation since November 2005. It contains four 2048$\times$2048 pixel HAWAII2-RG detectors and covers a total 20 arcmin$\times$ 20 arcmin field-of-view with a sampling of 0.3 arcsec per pixel (see Table \ref{tab:gal} for the angular sizes of the two galaxies). 

The observations of both galaxies were performed by sequentially shifting the object into the center of each detector for each new exposure. Additionally, a certain amount of dithering around the object was added to mitigate the effect of bad pixels. This chip-shifting method allows adjacent off-target exposures of the same detector to be used for sky subtraction as the two galaxies both have angular sizes comparable to the size of one detector. Pipeline processed images with sky subtraction (including dark subtraction, flatfielding with dome flat exposures, basic crosstalk removal, 2MASS absorption estimate and sky subtraction using neighboring images) are retrieved from the CFHT science data archive and the total number of exposures is listed in Table \ref{tab:file}. The actual pixel scale of the retrieved images is 0.304 arcsec per pixel. Each exposure is stored in a multi-extension fits file with four frames corresponding to the four different chips. As each exposure contains only one on-target frame, with the off-target frames used for background subtraction, we have a total of  99 narrowband and 112 broadband sky subtracted frames for NGC\,5055 and 61 narrowband and 61 broadband sky subtracted frames for NGC\,6946. 


In this paper we use ``narrowband map'' to refer to the observed line-plus-continuum frames obtained with the Br$\gamma$ filter (central wavelength 2166nm, bandwidth 29.5nm), ``Br$\gamma$ map'' to refer to the Br$\gamma$ emission line-only frames, and ``broadband map'' to refer to the observed continuum frames obtained with the Ks filter.

\subsubsection{NGC\,5055 Reduction}

For the observations of NGC\,5055, the different exposures from the same detector are well-aligned (as reduced by the CFHT pipeline), but not for exposures from different detectors. So we first combine all the images from the same detector into one single image using IRAF. A sigma clipping algorithm is used to eliminate spurious features, mostly the cross-talk from bright stars. All the sky subtracted CFHT images are given a `chipbias' to prevent 16-bit wrapping of negative numbers into high positive numbers and all the bad pixels are given a value of 0, so a global `sky' value for each image, the mode\footnote{Mode is the global peak of the value distribution. It is robust against the tail of positive sources to estimate the sky value.} value of the whole map, with values close to the `chipbias', is first subtracted from each image. This process reduces our collection of NGC\,5055 images from 99/112 (narrowband/broadband) to eight: one combined frame for each detector and each filter. The relative astrometry between the two bands of the same detector is stable so we use the broadband images to derive the alignment transformation between detectors and apply them to both the broadband and narrowband images. Thirty point-like sources (foreground stars) are chosen near and within the galaxy and the coordinates of their centers are determined. These coordinates are then used as input to the IRAF tasks `geomap' and `geotran' to align three of the images relative to the fourth one, with a manual inspection of the geometrical transformation process to mask out outliers due to mis-centering. A higher order fit is used to allow both offsets and rotation/stretching to be corrected. The same transformation is applied to align three of the narrowband images, relative to the fourth one. The four images from the four different detectors are then combined for the narrowband and the broadband separately.

The chip1 and chip4 Ks images of NGC\,5055 suffer from severe crosstalk from the bright guide star and thus only the chip2 and chip3 Ks images are used for this galaxy. This reduced the number of images combined to produce the Ks image to fifty-six 20s exposures. However, for the sole purpose of continuum subtraction of the Br$\gamma$ image, the resultant image still provides sufficient depth (see section \ref{calib}). 

\subsubsection{NGC\,6946 Reduction}

For this galaxy we had to proceed with a different approach than for NGC\,5055 as the different exposures, even from the same detector, is not well-aligned. No reason is noted on the archival images but possibly it is because of the large dither offset from one exposure to the other. Thus, for the NGC\,6946 images, we manually identified 137 point-like sources across the whole field, in order to align all images relative to each other and across the entire FOV, also correcting for distortions. The IRAF tasks `geomap' and `geotran' are again used, with manual inspection, to align the 61 images with reference to the first image for both the narrowband and the broadband separately. The rest of the approach remains the same as that of NGC\,5055 and in this case there is no problem with crosstalk from the guide star. Then the 61 images are combined using IRAF task `imcombine' for the two bands separately.

\subsubsection{Continuum Subtraction and Photometric Calibration}
\label{calib}

To obtain the Br$\gamma$ emission line image, we remove the underlying stellar continuum from the narrowband image. We use the broadband Ks image for this purpose. Both the narrowband image and the broadband image are first rescaled by the exposure time, and the Ks image resolution is matched to the Br$\gamma$ image resolution with the IRAF task `psfmatch', with a set of point--like sources ($\sim$10, these are presumably late type main sequence stars in our own Galaxy, and are not saturated in our images). The stars have FWHMs from 1.1$^{\prime\prime}$ to 1.4$^{\prime\prime}$, and the differences between the two bands are less than 0.2$^{\prime\prime}$. Then, photometry of the same set of stars for both NGC\,5055 and NGC\,6946, respectively, is performed in both bands; the average narrowband-to-broadband fractions are determined for both galaxies separately. The average fractions serve as a fiducial number we use to perform the continuum subtraction, assuming the Br$\gamma$ features from the stars are negligible. 

Many factors may actually affect the accuracy of this number, such as spectral type differences between stars and H{\sc ii} regions, differential foreground extinction, etc. 
 The Ks band targets mostly the Rayleigh-Jeans tail of stellar emission,
and, for our galaxies, we do not expect large in--band color variations
due to variations in stellar populations or due to the presence of
spatially--variable dust attenuation. Stellar population synthesis models
(Starburst99) indicate that variations in the stellar flux ratio between
the two wavelength extremes of the Ks filter are less
than 3\% for an age range between 6 Myr and 10 Gyr. We verify the impact
of in--band dust attenuation variations by considering our largest
A$_V$=3.5~mag derived from H$\alpha$/Br$\gamma$. This corresponds to
A$_{Ks}\sim$0.18~mag, when including the factor $\sim$2 reduction in
attenuation for the stellar continuum relative to the ionized gas \citep{calzetti2000, hao2011}.
 The variations in stellar color across the
Ks filter due to this value of A$_{Ks}$ is about 1\%. 
We also cross-check these scaling numbers against an optical continuum subtraction method \citep{hong2011}, which analyzes the statistics of the continuum subtracted images with different continuum fractions and gives a robust estimate of the acceptable continuum fraction range. The acceptable range we get from this method includes our estimate from rescaling star fluxes, which shows that the two methods agree. Having performed this consistency check, we ultimately use the factor determined from the stars and inspect the continuum subtracted Br$\gamma$ images to make sure the subtraction looks reasonable. The fractions of the broadband fluxes adopted are 0.0890 and 0.0864 for NGC\,5055 and NGC\,6946 respectively. The contribution of Br$\gamma$ lines to the Ks filter (FWHM 315nm) is negligible ($\le$1.5\% even for the youngest star forming regions), thus it is not necessary to correct for the over-subtraction of the Br$\gamma$ line itself when subtracting the continuum. In addition to the Br$\gamma$ line, the H$_2$ rotational transitions may contribute to the emission in the Ks filter. These, however, tend to contribute at most 2.4\%, and more typically less than 1\%, to the total Ks band, even in active starburst systems \citep{calzetti1997,rosenberg2013}.

Despite the care taken in the continuum subtraction, we still test for the potential effect of incorrect continuum subtraction. For larger the determined value of continuum subtraction fraction, we find the smallest multiplicative value that gives over-subtracted Br$\gamma$ images for the two galaxies; this value is taken as the upper bound of possible continuum subtraction fractions. The symmetric value on the smaller side relative to this is taken as the lower bound, as it is more difficult to determine the point where under-subtraction occurs. The adopted ranges are 0.0880 to 0.0900 and 0.0852 to 0.0876 for NGC\,5055 and NGC\,6946 respectively. A set of aperture photometry measurements on H{\sc ii} regions are performed on images with different continuum subtraction levels (the optimal value and the two boundaries determined above), which reveals an average of 10\%-15\% variation in flux measurement between the optimal subtraction fraction and the boundaries. This variation depends on how bright the sources are and the difference could rise to nearly a factor of 2 for the extremely faint H{\sc ii} regions, with low signal-to-noise (S/N$\lesssim$1). We thus adopt a S/N$>$3 cut that immediately excludes such extreme faint regions (see section \ref{error}). The central parts of both galaxies and the centers of the bright stars show artifacts on the line-only Br$\gamma$ maps, mainly due to saturation of these regions. The artifacts may also be enhanced by small misalignments in the final combined images, because the centroids are poorly determined in saturated sources. We make no use of the bright stars and try to avoid the regions (central $\sim$10$^{\prime\prime}$) affected by these artifacts in our analysis. Also, because NGC\,6946 is in a very crowded field, the foreground stars with obvious residual after continuum subtraction are masked out. The final Br$\gamma$ images for NGC\,5055 and NGC\,6946 both show an uneven background, possibly due to an imperfect sky subtraction in the pipeline and the interpolation in the process of aligning the different exposures. To flatten the background, we use the IRAF package `imsurfit' to fit a 2-dimensional low-order surface and subtract the background from the images. 
Although this approach is effective at flattening the line-only images, it may not be sufficient for large scale analysis (the peak-to-peak difference in the background for the images matched to the resolution of the IR data is about 5~$\sigma$). We thus avoid deriving integrated luminosities for the whole galaxy or radial trends, and limit our analysis to the typical (or somewhat larger) sizes of H{\sc ii} regions.

The Br$\gamma$ maps are calibrated with the header keyword `phot\_c0', which is determined using four well modeled spectrophotometric standards (CALSPEC models) and filter transmission curves, with the information from the WIRCam site\footnote{http://www.cfht.hawaii.edu/Instruments/Imaging/WIRCam/WIRCamStandardstars.html}$^{,}$\footnote{http://www.cfht.hawaii.edu/Instruments/Imaging/WIRCam/dietWIRCam.html\#UI}$^{,}$\footnote{http://www.cfht.hawaii.edu/Instruments/Filters/wircam.html}. A comparison between this calibration and the  {\it HST} P$\alpha$ fluxes for these two galaxies, assuming Case-B \citep{osterbrock1989} recombination conditions and similar extinction at P$\alpha$ (1.8756~$\mu$m) and Br$\gamma$ (2.166~$\mu$m), shows consistency within 5\%. The uncertainty of `phot\_c0' is around 0.1 magnitude ($\sim$10\%). Including the uncertainty from continuum subtraction, we expect about a 15\% uncertainty in the absolute flux calibration  (sum in quadrature of 10\% from the calibration keyword uncertainty and about 10\% from the continuum subtraction uncertainty), which could be less for bright H{\sc ii} regions (close to $\sim$10\%) as the continuum subtraction mainly affects the fainter regions.  The final continuum subtracted images have 1-$\sigma$ noise level of 2.12$\times10^{-18}$ ergs$\cdot$s$^{-1}\cdot$cm$^{-2}\cdot$pixel$^{-1}$ (0.3 arcsec per pixel) for NGC\,5055 and 2.48$\times10^{-18}$ ergs$\cdot$s$^{-1}\cdot$cm$^{-2}\cdot$pixel$^{-1}$ for NGC\,6946. 

\subsection{Herschel PACS Data}
\label{herschel}

Herschel PACS observations in all three bands (70, 100 and 160~$\mu$m) are available for both galaxies from the KINGFISH project. The observational program and data processing procedures for KINGFISH are described in detail in \citet{kennicutt2011,dale2012}. Briefly, PACS imaging was obtained with 15$^\prime$-long scan maps, along two perpendicular axes for improved image reconstruction, at the medium scan speed of 20$^{\prime\prime}$s$^{-1}$. The 45$^\circ$ orientation of the array with respect to the scan direction contributes to a more uniform spatial coverage. The raw (`Level 0') data were processed using Version 5 of HIPE \citep{ott2010}. Besides the standard pipeline procedures, the conversion from Level 0 to Level 1 data included second level deglitching and corrections for any offsets in the detector sub-matrices. Scanamorphos\footnote{http://www2.iap.fr/users/roussel/herschel} \citep{roussel2012} was used to process the Level 1 PACS scan map data to Level 2 final products, where the pixel size is $\sim$ one-fourth of the beam FWHM, i.e., 1.4$^{\prime\prime}$ at 70~$\mu$m, 1.7$^{\prime\prime}$ at 100~$\mu$m, and 2.85$^{\prime\prime}$ at 160~$\mu$m. A multiplicative factor was also implemented to update the PACS calibration, which is equivalent to the version 26 calibration\footnote{http://herschel.esac.esa.int/twiki/bin/view/Public/PacsCalTreeHistory?template=viewprint}.  We adopt the latest values of 10\% for all three PACS bands as absolute calibration uncertainty for our extended sources\footnote{https://nhscsci.ipac.caltech.edu/pacs/docs/Photometer/PICC-NHSC-TR-034.pdf}$^{,}$\footnote{http://herschel.esac.esa.int/twiki/pub/Public/PacsCalibrationWeb/ExtSrcPhotom.pdf}.  The data used in this paper are processed with Scanamorphos version 12.5 while the most recent KINGFISH data are processed with Scanamorphos version 16.9; checks performed against the versions 16.9 of the {\it Herschel} maps show $<$1\% difference in photometry. 

\subsection{H$\alpha$ Data}
\label{Ha}

\begin{figure*}[h]
\center{
\includegraphics[scale=0.8]{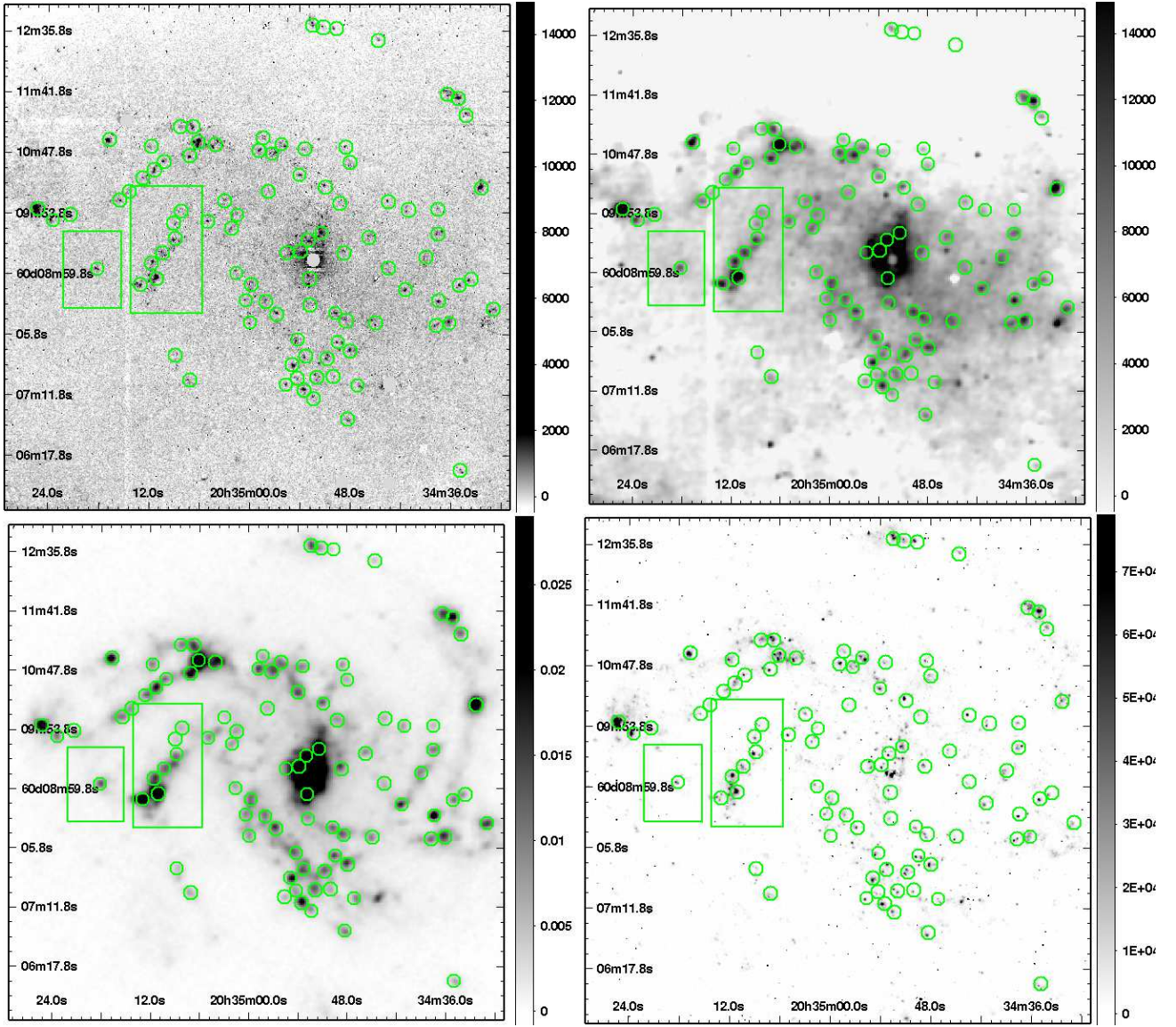}
}
\caption[]{\label{fig:6_7} From top left to bottom right: Br$\gamma$, Br$\gamma$ at 70~$\mu$m resolution, 70~$\mu$m and H$\alpha$ images for NGC\,6946. The green circles are the apertures we adopted for our analysis at 70~$\mu$m resolution. The boxes are the local regions, with relatively even background, selected for the comparison between the two methods of subtracting the local background (Section \ref{aper}). North is up and east is left.  The image size is about 450$^{\prime\prime}\times$450$^{\prime\prime}$ (about 14.8$\times$14.8 kpc) for each panel. The uneven background of the Br$\gamma$ images is greatly enhanced after convolving to the 70~$\mu$m resolution. The units for the grayscale bars are, 10$^{-20}$ ergs$\cdot$s$^{-1}\cdot$cm$^{-2}$ per pixel for Br$\gamma$ and H$\alpha$ with pixel sizes of 0.304$^{\prime\prime}$ and for Br$\gamma$ at 70~$\mu$m resolution with a pixel size of 1.4$^{\prime\prime}$, and Jansky (Jy) per pixel for 70~$\mu$m with a pixel size of 1.4$^{\prime\prime}$.}
\end{figure*}

\begin{figure*}[h]
\center{
\includegraphics[scale=0.8]{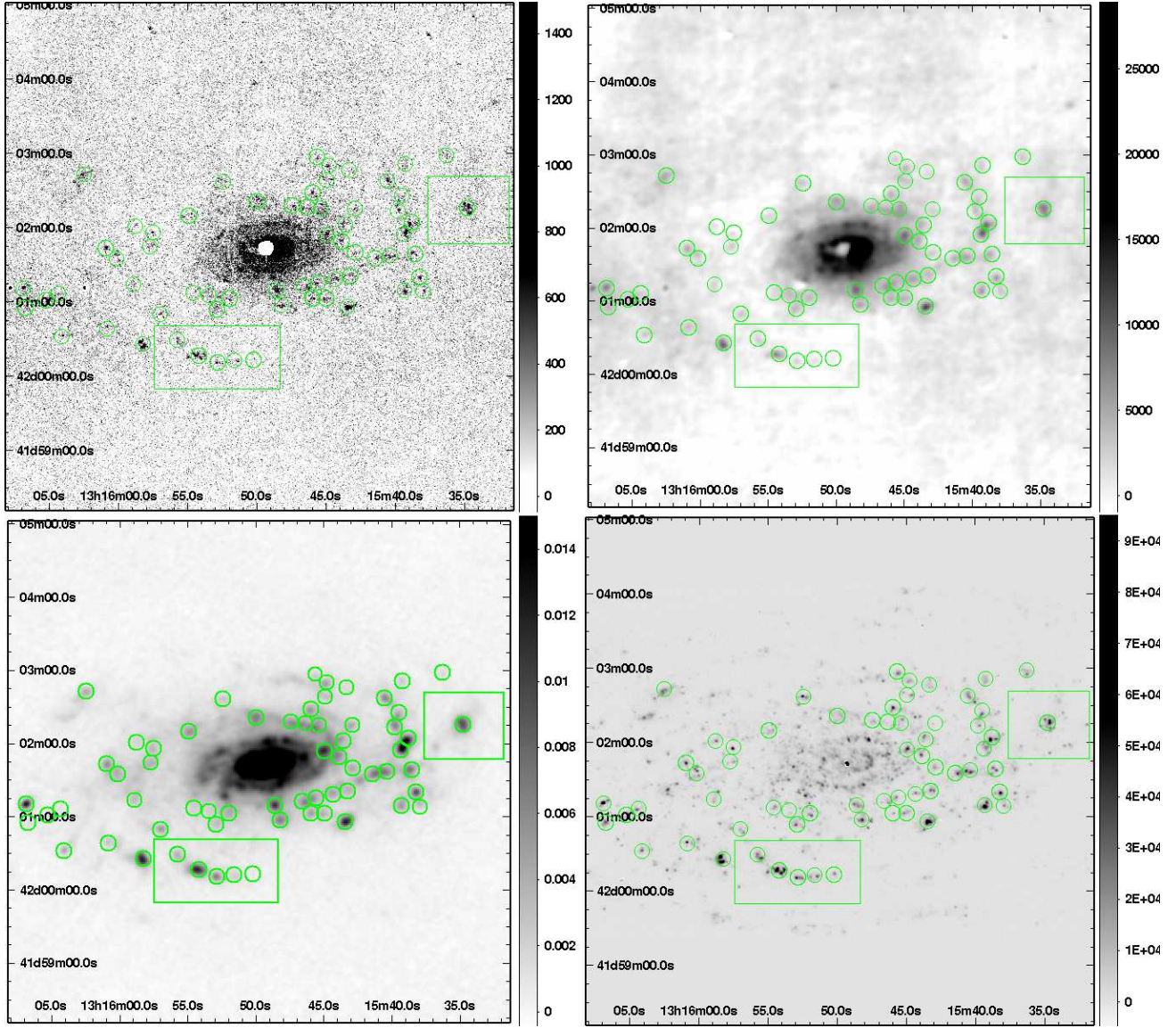}
}
\caption[]{\label{fig:5_7} From top left to bottom right: Br$\gamma$, Br$\gamma$ at 70~$\mu$m resolution, 70~$\mu$m and H$\alpha$ images for NGC\,5055. The symbols and units for the grayscale bars are the same as described in Fig. \ref{fig:6_7}. North is up and east is left. The image size is about 410$^{\prime\prime}\times$410$^{\prime\prime}$ (about 15.8$\times$15.8 kpc) for each panel. The uneven background of the Br$\gamma$ images is greatly enhanced after convolving to the 70~$\mu$m resolution.}
\end{figure*}

To correct for the small amount of dust extinction in our Br$\gamma$ maps, we use the H$\alpha$ images of NGC\,6946 from the Spitzer Infrared Nearby Galaxies Survey \citep[SINGS,][]{kennicutt2003} and of NGC\,5055 from the Local Volume Legacy \citep[LVL,][]{kennicutt2008,dale2009,lee2011} survey (Fig. \ref{fig:6_7} \&\ref{fig:5_7}). The H$\alpha$ photometry is corrected for [NII] contamination using the [NII]/H$\alpha$ ratios listed in \citet[][and the references therein]{kennicutt2009} (and also in Table \ref{tab:gal}). The SINGS H$\alpha$ map of NGC\,6946 only covers the central $\sim$15~kpc by 15~kpc  (about 450$^{\prime\prime}$ by 450$^{\prime\prime}$) area of the galaxy (Fig. \ref{fig:6_7}), and we limit our analysis to this region. We use the H$\alpha$ map to correct for dust attenuation and the typical attenuation corrections of H{\sc ii} regions are A(H$\alpha$)$\sim$1.65 mag for NGC\,5055 and A(H$\alpha$)$\sim$1.12 mag for NGC\,6946 (see section \ref{extinction} and Table \ref{tab:gal}), after removal of foreground Galactic extinction, as derived using the method described in the following section. 

\subsubsection{ Extinction Correction}
\label{extinction}

With a simple Case B recombination assumption, the ratio of H$\alpha$ emission to Br$\gamma$ emission from the same H{\sc ii} region is expected to be $\sim$103 \citep{osterbrock1989} for an H{\sc ii} region with electron temperature of 10,000 K and electron density of 10$^2$ cm$^{-3}$, changing by only 35\% for electron temperatures in the range 5,000--20,000 K and by only 2\%--3\% for electron densities in the range 10$^2$--10$^4$ cm$^{-3}$. Figure \ref{fig:HavB} shows the ratio of $\rm\Sigma(H\alpha,obs)$ to $\rm\Sigma(Br\gamma,obs)$ as a function of $\rm\Sigma(70)$, where the data should lie about the y=103 line if there were no extinction. In general, the data for both galaxies show a weak trend for decreasing ratio, thus increasing extinction,  for increasing $\rm\Sigma(70)$. The ratio is mostly below 103, showing that extinction is present. Although a couple of regions in NGC\,6946 have a ratio greater than 103, they all have significantly larger error bars indicating lower S/N in at least one of the two line measurements.

We use the deviation of the line ratio from a value of 103 to estimate the extinction of both bands. Since $\rm A(H\alpha)=2.54\cdot E(B-V)$ and $\rm A(Br\gamma)=0.36\cdot E(B-V)$, then we can get the nebular gas color excess as
\begin{equation}
\label{equ:ebv}
\rm E(B-V)=\frac{log(103\frac{\Sigma(Br\gamma,obs)}{\Sigma(H\alpha,obs)})}{0.87},
\end{equation}
where $\rm\Sigma(Br\gamma,obs)/\Sigma(H\alpha,obs)$ is the observed surface luminosity ratio. 
We calculate the color excess for each region and apply the extinction correction to $\rm\Sigma(Br\gamma)$. The color excess is calculated under the assumption of foreground dust extinction \citep{calzetti2000}, because the dust extinction corrections in the NIR regime are relatively insensitive to the dust geometry \citep{calzetti1996}. The extinction corrected $\rm\Sigma(Br\gamma)$ is then used as a reference SFR indicator. The median E(B-V) values for our regions of NGC\,5055 and NGC\,6946 at 70~$\mu$m resolution are listed in Table \ref{tab:gal}, and correspond to a median of 23.6\% and 15.7\% correction of the Br$\gamma$ luminosity for NGC\,5055 and NGC\,6946, respectively. 


H{\sc ii} regions in metal--rich galaxies like NGC\,5055 and NGC\,6946 often have measured electron
temperatures that are slightly lower than our default 10,000~K. A temperature of 8,000~K
is more common \citep{bresolin2004}, which yields an H$\alpha$ to Br$\gamma$ ratio of 99.3. For this case, the derived E(B-V) from equation (\ref{equ:ebv}) would be about 5\% lower and the extinction correction about 1\% lower for the Br$\gamma$. Given that the conversion from Br$\gamma$ luminosity to SFR(Br$\gamma$) through H$\alpha$ luminosity would also be about 4\% lower (99.3/103=0.964), the SFR(Br$\gamma$) will be about 5\% lower and thus all the derived parameters which are proportional to SFR(Br$\gamma$) in the rest of the paper would be about 5\% lower. Since this does not have a major impact, we keep our default H$\alpha$ to Br$\gamma$ ratio of 103 corresponding to an electron temperature of 10,000 K, for generality.


\section{Analysis}

\subsection{Robustness of the Br$\gamma$ Data}

In order to evaluate our Br$\gamma$ data reduction and calibration, we
derive luminosity functions for the H{\sc ii} regions in the two galaxies (section \ref{LF}). We
also compare a small subset of bright H{\sc ii} regions in NGC6946 with
published radio free-free measurements (section \ref{vsFF}).

\subsubsection{Luminosity Functions}
\label{LF}

\begin{figure*}[b]
\center{
\includegraphics[scale=0.8]{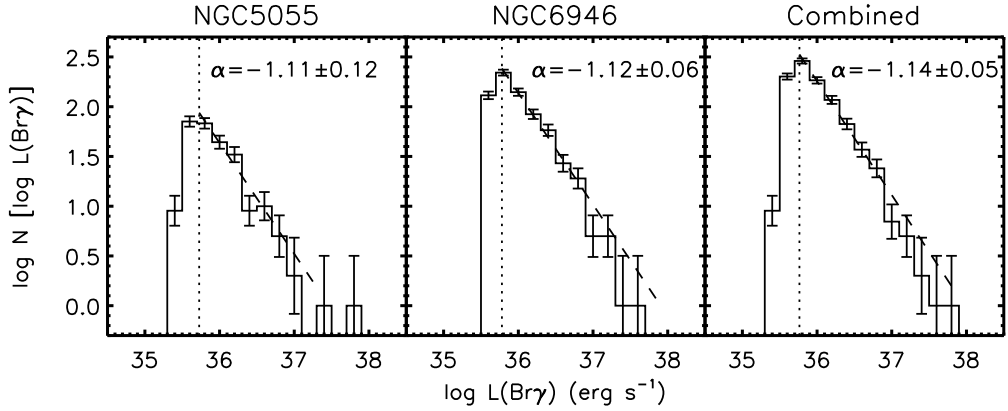}
}
\caption[]{\label{fig:LF} Observed Br$\gamma$ emission luminosity functions. The vertical dotted lines are 5-$\sigma$ detection limit, logL(Br$\gamma$)$\sim$35.7. The dashed lines are power law fits, $\rm dN/dlogL(Br\gamma)\propto L(Br\gamma)^{\alpha}$, for the luminosity functions from the 5-$\sigma$ detection limit to the continuous bright end, which are all consistent with dN/dL(Br$\gamma$)$\propto$L(Br$\gamma$)$^{-2}$ or dN/dlogL(Br$\gamma$)$\propto$L(Br$\gamma$)$^{-1}$.}
\end{figure*}

The luminosity functions of the H{\sc ii} regions in both galaxies are produced from the Br$\gamma$ (uncorrected for extinction) images using the IDL program H{\sc ii}phot  \citep{thilker2000,thilker2002}. H{\sc ii}phot has an automatic algorithm to identify H{\sc ii} region seeds and then grow the H{\sc ii} regions from these seeds until it meets another H{\sc ii} region or reaches a given emission measure gradient. Background subtractions are applied to the identified H{\sc ii} regions, and both luminosity and S/N are measured. The luminosity function of the H{\sc ii} regions detected above 5~$\sigma$ are shown for both galaxies, individually and combined, in Figure \ref{fig:LF}. The luminosity functions all have a power law fit, $\rm dN/dlogL(Br\gamma) \propto L(Br\gamma)^{\alpha}$, with power law slope $\alpha$ consistent with -1 (shown on Fig. \ref{fig:LF}), which corresponds to a luminosity function, dN/dL(Br$\gamma)\propto$L(Br$\gamma)^{\alpha-1}$ and agrees with previous studies of H$\alpha$ luminosity functions \citep[e.g.][]{kennicutt1989,thilker2000,thilker2002}. The most luminous regions have L(Br$\gamma$) $\sim$ 10$^{37.7}$ ergs$\cdot$s$^{-1}$, corresponding to L(H$\alpha$) $\sim$ 10$^{39.7}$ ergs$\cdot$s$^{-1}$ (if no extinction). These values are also in agreement with previous findings \citep{thilker2000,thilker2002}. This provides a consistency check that our continuum subtraction and absolute photometric calibration are reasonable. The H{\sc ii} region luminosity functions have a 5-$\sigma$ detection limit of $\rm log[L(Br\gamma)]\sim35.7$, corresponding to star clusters with mass $\rm\sim4-7\times10^3\ M_\odot$ \citep[STARBURST99,][] {leitherer1999}, assuming an average age of 4-5 Myr and a fully sampled \citet{kroupa2001} stellar IMF. Using the estimates of \citet{cervino2002} adapted to our IMF, we expect fluctuations on the ionizing photon rate $\delta Q(H^{0})/Q(H^{0})\lesssim30\%$. Thus, our regions encompass a sufficiently large number of young stars that random fluctuations in the ionizing photon flux are not a concern.

\subsubsection{Comparison between Br$\gamma$ and Free-free Emission}
\label{vsFF}

\begin{figure*}[h]
\center{
\includegraphics[scale=0.8]{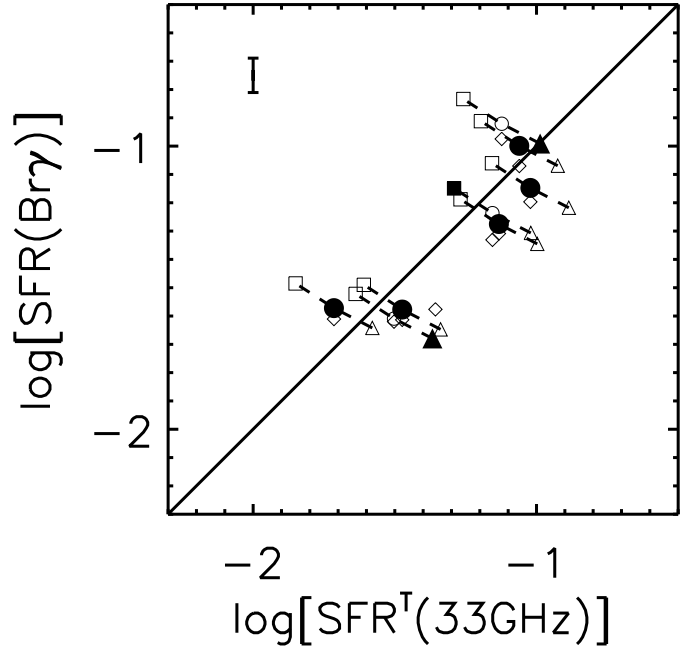}  
}
\caption[]{\label{fig:vsGBT} SFR(Br$\gamma$) as a function of SFR derived from thermal free-free emission at 33GHz \citep{murphy2011}.The triangles, circles and squares are values derived using electron temperatures of 5000, 10000 and 20000~K, respectively. The open diamonds are values calculated using Br$\gamma$ emission without internal extinction correction, assuming a typical electron temperature of 10000K.  The filled symbols show the preferred temperature for each region as derived in \citep{murphy2011}. The one open diamond in the middle of the plot without a corresponding filled symbol is for Enuc 4, which we have no coverage with our H$\alpha$ map thus no extinction corrected value. The typical error bar for SFR(Br$\gamma$) is shown on the upper left corner. The unit is $\rm M_{\odot}\cdot yr^{-1}$ for SFR.}
\end{figure*}

\begin{figure*}[h]
\center{
\includegraphics[scale=0.8]{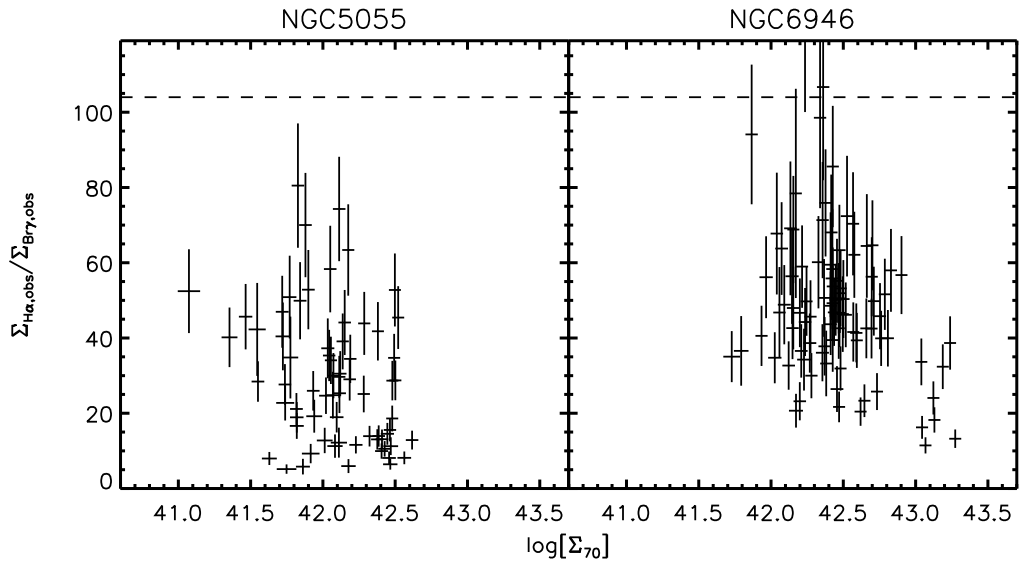}
}
\caption[]{\label{fig:HavB} $\rm\Sigma(H\alpha,obs)/\Sigma(Br\gamma,obs)$ as a function of $\rm\Sigma(70)$. The two panels are data for NGC\,5055 and NGC\,6946 respectively, after removal of foreground Galactic extinction. The dashed line indicates the extinction free value of $\rm\Sigma(H\alpha,obs)/\Sigma(Br\gamma,obs)$ ($\sim103$). Error bars are shown for each data point. The unit is $\rm ergs\cdot s^{-1}\cdot kpc^{-2}$ for for the luminosity surface density.}
\end{figure*}

To further test our Br$\gamma$ data, we compare the Br$\gamma$ emission of NGC\,6946 with the thermal free-free component at 33 GHz emission, as observed with the GBT \citep[][Fig. \ref{fig:vsGBT}]{murphy2011}. \citet{murphy2011} performed the spectral decomposition of the radio and far--infrared emission of the targeted regions, separating the free-free emission from other emission components. The GBT observations cover 10 pointings of NGC\,6946 (Fig. 1 in \citealt{murphy2011}) and have a FWHM about 25$^{\prime\prime}$. Our Br$\gamma$ map of NGC\,6946 has artifacts in the galaxy center, so we only compare the extranuclear measurements. For our Br$\gamma$ emission, we also use an H$\alpha$ map of NGC\,6946 for internal extinction correction (see section \ref{Ha} \& \ref{extinction}), while the radio emission is insensitive to dust attenuation. Both images are convolved with a gaussian of FWHM 25$^{\prime\prime}$ to match the resolution of the GBT 33GHz observations. The Enuc 4 \citep[the number four extranuclear region as named by][]{murphy2011} is the only region among the 10 observed to have statistically significant excess of anomalous microwave emission in the 33 GHz band
\citep{murphy2010}; however, this region is located outside the coverage
of our H$\alpha$ map, implying that we do not have extinction--corrected
Br$\gamma$ flux for this region. Thus, our comparison between the
Br$\gamma$ and the free-free emission uses a total of eight regions.
 A test with  the extinction correction for the regions before and after the convolution shows less than 5\% difference. As our analysis is on a region by region basis, we adopt the extinction correction after the convolution for our Br$\gamma$ measurements. Photometry is measured with apertures and background annuli similar to those used by \citet{murphy2011}. 

The galaxy-wide star formation rate calibrations do not apply precisely to a single-age young populations like those in H{\sc ii} regions. Here, our apertures correspond to regions consisting of a few H{\sc ii} regions, and both the free-free and the Br$\gamma$ measurements are actually reflections of the ionizing photons, Q(H$^0$), from the young stars. However, for a more direct comparison with previous results, we still calculate the SFR(Br$\gamma$) using the estimated Q(H$^{0}$) flux \citep{osterbrock1989}, with extinction corrected Br$\gamma$. SFR$^{T}$(33GHz), the star formation rate derived from the thermal free-free component of the 33 GHz radio emission, is calculated using Equation (11) from \citet{murphy2011}. Both quantities are calculated assuming electron temperatures of 5000K (triangles), 10000K (circles) and 20000K (squares) to account for the range of variations in electron temperatures found in H{\sc ii} regions, which depend on galacto-centric radius and metallicity \citep{shaver1983}. The electron temperature directly affects the intrinsic H$\alpha$/Br$\gamma$ ratio \citep{osterbrock1989}, thus affecting the extinction correction for the Br$\gamma$ emission using the H$\alpha$ emission. It also affects the conversion both from measured luminosities of Br$\gamma$ and 33GHz to Q(H$^{0}$) flux and from Q(H$^{0}$) flux to SFR, as shown in Equation (11) from \citet{murphy2011}. Values calculated using Br$\gamma$ emission without internal extinction correction, assuming a typical electron temperature of 10000K, are shown as open diamonds, representing about 3\%-18\% corrections for the filled circles in the Br$\gamma$ emission (with a median $\sim$12\%) after removal of foreground Galactic extinction.  The two inferred SFRs show about 27\% difference on average, for the preferred temperature as shown with the filled symbols. Even when the extreme variations in electron temperature (from 5000K to 20000K) are taken into account, the two inferred SFRs are still in good agreement. This provides again a consistency check that our data reduction and calibration are reliable. Furthermore, our comparison shows that the Br$\gamma$ data, once corrected for dust extinction using the H$\alpha$, do not appear to require additional dust corrections, implying that the fraction of ionizing photons completely obscured by dust at 2.16~$\mu$m is negligible.

 Our 3-$\sigma$ detection limit within the 12.5$^{\prime\prime}$ radius aperture is SFR(Br$\gamma$)$\sim$0.0017, or about one order of magnitude fainter than the faintest region analyzed by \citet{murphy2011}. The higher sensitivity of Br$\gamma$ imaging suggests that this
may become a preferred approach to derive a dust--free SFR indicator over extended sections of star-forming galaxies.

\subsection{PSF-Matching}
In order to measure the photometry consistently among the different maps (H$\alpha$, Br$\gamma$ and 70~$\mu$m), we first need to convolve both the H$\alpha$ and the Br$\gamma$ maps to the resolution of the 70~$\mu$m map ($\sim$ 5.5$^{\prime\prime}$). The point-spread function (PSF) FWHM of the Br$\gamma$ is about 1.1$^{\prime\prime}$ and the H$\alpha$ only slightly better (about 1$^{\prime\prime}$). 

For the PACS PSFs, we choose the observed PSFs of the asteroid Vesta available on the PACS website\footnote{http://herschel.esac.esa.int/twiki/bin/view/Public/PacsCalibrationWeb}. We use the PSF with the same scan speed (20$^{\prime\prime}$s$^{-1}$) and similar orientation (42 deg) to the KINGFISH observations. We measure the curve of growth of the observed PSF and find that more than 2/3 of total flux is included within a radius of 6$^{\prime\prime}$, slightly larger than the PSF FWHMs, 5.5$^{\prime\prime}$. Although we think this specific observed PSF should be the best match to our observations, the curve of growth shows less than 1\% difference among observed PSFs with different observational parameters. The model PSFs show a different curve of growth but they still contain more than 2/3 of total flux within the radius of 6$^{\prime\prime}$. Thus we choose 6$^{\prime\prime}$ as our aperture radius for the 70~$\mu$m resolution. The PACS PSFs are not spherically symmetric and are dependent on the scan direction, thus the PSFs are rotated by the scan angle (relative to north, as our maps are reprojected with north-up) of our PACS maps. A visual comparison of the rotated PSF and the brightest point-like sources on the FIR maps of both galaxies shows differences of only a few degrees of rotation. A comparison between the photometry of Br$\gamma$ maps convolved with one PSF and the same PSF rotated by 60 degrees shows only about 7\% difference.

Finally, the H$\alpha$ and Br$\gamma$ maps are convolved with the observed PSF, rotated with the scan angle of the {\it Herschel} observation for each galaxy, to produce the H$\alpha$ and Br$\gamma$ maps at 70~$\mu$m resolution. The convolved H$\alpha$ and Br$\gamma$ maps are then registered to the same pixel sizes as the 70~$\mu$m maps with astrometry matched using the brightest sources in the galaxies.  Although the PSF FWHM of the Br$\gamma$ is small compared to the 70~$\mu$m, we still convolve the 70~$\mu$m images with the extracted PSF from several unsaturated bright stars in the CFHT narrowband images to better match the resolution between the two. We are confident that the convolution brings the optical, NIR and 70~$\mu$m maps to similar resolution and adds a photometric uncertainty no larger than 2\%, combining both the difference in observed PSFs with different observation parameters and possible small mismatch of PSF rotation angle. The final PSF FWHM for the 70~$\mu$m images, the Br$\gamma$ images and the H$\alpha$ images is $\sim$5.6$^{\prime\prime}$.

\subsection{Photometry in Selected Regions}
\label{photometry}

\subsubsection{Aperture Photometry}
\label{aper}

Star forming regions are selected on the Br$\gamma$ images at the original
resolution, by placing apertures with radii of 6$^{\prime\prime}$ centered on the peaks of ionized gas emission.
Care is taken so that apertures do not significantly ($<$5\%) overlap. Further inspection of the 70~$\mu$m maps and the degraded Br$\gamma$ and H$\alpha$ maps is performed to make sure that each aperture covers at least one 70~$\mu$m emission peak and is as close as possible to the emission peak both at 70~$\mu$m and in the emission line images. Sources near the continuum subtraction residuals of bright stars and the galaxy center are not included as to avoid possible contamination from the residuals. The central region of NGC\,5055 is identified to have an AGN by \citet{moustakas2010}, and it shows one single bright source in the center covering up a region of about 10$^{\prime\prime}$ radius ($\sim$ 400pc). This region is also avoided when selecting apertures. We selected a total of 73 regions for NGC\,5055 and 94 regions for NGC\,6946 (Fig. \ref{fig:6_7} \& \ref{fig:5_7}). We note that these regions are not the same as those described in
section \ref{LF}; H{\sc ii}phot could not be effectively applied to our resolution degraded images, and we resorted to a manual selection. In addition, due to the lower resolution, the regions used for the bulk of our present analysis are larger (and more luminous) than those used for deriving the luminosity functions.

Following the studies of \citet{calzetti2005}, \citet{li2010} and \citet{liu2011}, local background subtraction is essential for photometric measurements in spatially resolved regions in order to minimize the effect of diffuse IR emission that is not related to current star formation activity. Furthermore, it will serve to mitigate the effects of the uneven background in our Br$\gamma$ maps. We perform the local background subtraction with two methods: one is to identify a local background region for different apertures \citep{calzetti2007,li2010}; the other is to use an annulus outside each aperture not much larger than the aperture radius itself, a method that follows standard practice for photometric measurements. For the first method, we choose two local background regions within each of the two galaxies, with relatively even background on the Br$\gamma$ map and similar physical conditions, i.e. on the same spiral arm, to ensure that the diffuse IR emission is similar, by visual inspection. For the second method, we adopt an annulus whose inner radius and outer radius are two times and four times the aperture radius, respectively. The width is chosen to be larger than two times the PSF FWHM and is outside of the second Airy ring. Each annulus is thus close enough to its aperture to minimize the effects of the uneven background while still containing enough pixels for a good local background determination and not including significant emission from the region itself. For both methods, the mode is used for the background level due to crowding in some regions. The background levels between the two methods agree with each other with differences of less than several percent when comparisons are performed in uncrowded regions. The results prove that both methods are valid for effectively removing the local background in our maps. Due to the uneven background in our Br$\gamma$ maps, we choose the annulus method, which is more robust against such variations. 

The 70~$\mu$m resolution aperture correction from a 6$^{\prime\prime}$ radius to infinity is 1.466, which is determined using the PACS observed PSFs convolved with the Br$\gamma$ PSF. The local background subtraction method described above is included in the aperture correction. We use aperture corrections to best recover the true luminosity of each source we select, but they do not affect the analysis, as all the images are degraded to the same resolution. The aperture corrections are only perfectly valid for point sources; however, our sources are usually compact and the underlying extended emission is relatively weak, thus this correction provides a reasonable estimate of the true luminosity of our sources.

\subsubsection{Error Estimate and Signal-to-noise Criteria}
\label{error}

We correct the H$\alpha$ and Br$\gamma$ photometry  for foreground Galactic extinction \citep{schlegel1998, odonnell1994} with a single value (Table \ref{tab:gal}), which is significant for NGC\,6946 as it is close to the Galactic plane, while the Galactic extinction for the FIR bands is negligible. This correction could be a potential source of systematic error if the adopted values are not correct. 

The major uncertainty in the photometry is the calibration uncertainty: 10\% at 70~$\mu$m, 15\% at Br$\gamma$, combining both the absolute calibration uncertainty and the continuum subtraction uncertainty, and 10\% at H$\alpha$ (see Section \ref{data}). The uncertainty in the local background level is also included. We take a 2\% uncertainty in the convolution with the observed PSF at different rotation angles for the three bands. Other uncertainties, such as the global background uncertainty, are relatively small. The final photometric uncertainty is usually dominated by the calibration uncertainty; the uncertainty in local background estimate affects the faint regions more than brighter regions.

We accept as `good data' regions with S/N$>$3 for all the three images at 70~$\mu$m, H$\alpha$ and Br$\gamma$. The noise is determined using the variation around the mode in the local background annuli of each region. This S/N cut removes the faint regions whose Br$\gamma$ photometry is complicated by the uneven background and uncertainties in the continuum subtraction. 
Our S/N$>$3 cut corresponds to different
luminosity cut-offs for different regions, as some of the regions fall in high background
noise areas. In our `good data', only five regions have S/N$<$5, which are regions in the East
and West outskirts of the galaxies (Figures \ref{fig:5_7} and \ref{fig:6_7}).
About 77\% of the regions have S/N greater than 10. The least luminous region has 
a L(Br$\gamma$)=10$^{36.6}$~erg$\cdot$s$^{-1}$ with S/N=10.7, corresponding to a S/N=3 limit
 of L(Br$\gamma$)=10$^{36.1}$~erg$\cdot$s$^{-1}$ in the low background noise area. The lowest
S/N region (S/N=3.5) has a L(Br$\gamma$)=10$^{36.7}$~erg$\cdot$s$^{-1}$, corresponding to
a S/N=3 limit of L(Br$\gamma$)=10$^{36.6}$~erg$\cdot$s$^{-1}$ in the high background noise area.
We note that the minimum L(Br$\gamma$)= 10$^{36.6}$~erg$\cdot$s$^{-1}$ is about eight times
higher than the limit discussed in section \ref{LF} for the Br$\gamma$ luminosity
function. This difference is explained by the larger apertures
we use in this analysis relative to those of section \ref{LF}, which causes blending
of H{\sc ii} regions. The larger apertures are dictated by the larger 70 $\mu$m PSF
than the native Br$\gamma$ PSF. Thus, the same argument about
the random sampling of the stellar IMF not being an issue when deriving SFRs
applies to our overall analysis, as well. Comparing the luminosity distribution
of our selected regions with the luminosity functions derived using H{\sc ii}phot (see
section \ref{LF}), we find a 50\% completeness limit at L(Br$\gamma$)$\sim$10$^{37}$~erg$\cdot$s$^{-1}$.
80\% of the regions we use for the L(70)--SFR analysis have luminosity brighter
than this limit, implying that our results are robust relative to completeness
issues.
For the `good data', we have median uncertainties (and uncertainty ranges) of about 17\% (15-33\%) in Br$\gamma$, 11\% (10-30\%) in H$\alpha$, 10\% (10-19\%) in 70~$\mu$m. The final number of regions used in our analysis is 68 for NGC\,5055, and 92 for NGC\,6946.

\subsubsection{Other Sources of Bias and Uncertainty}
\label{other}

Absorption of Lyman continuum photons (LyC) by dust is a potential source of bias that decreases the luminosity of hydrogen recombination lines and of the free-free continuum. Estimating this effect observationally is challenging, and models are usually used to estimate the impact of this effect. We follow the parametrization of \citet{dopita2003}, where the lost fraction of H$\beta$ luminosity to LyC absorption by dust is modeled as the product of the oxygen abundance and ionization parameter. For the typical oxygen abundance (about solar) and ionization parameter range (U$\sim$10$^{-3}$) expected in our regions, we estimate the LyC absorption by dust to be at the level of 10\% or less. \citet{hunt2009, draine2011} suggests similar scenarios except in very extreme conditions. We thus neglect this small negative contribution in our analysis.

Leakage of ionizing photons out of the regions is also a potential source of bias. Many studies \citep[e.g.][]{ferguson1996, thilker2002, oey2007} place the fraction of ionizing photons escaping from H{\sc ii} regions at the 40\%--60\% level, as evaluated from the {\em observed} H$\alpha$. When accounting for the differential dust attenuation between H{\sc ii} regions and the diffuse medium, the fraction of diffuse H$\alpha$ is reduced by roughly a factor 2 \citep{crocker2012}. We thus expect that our Br$\gamma$ measurements will be systematically biased low compared to the number of ionizing photons produced by about 20-30\%. The dependence of this bias on the region's characteristics (luminosity, density, etc.) is still under investigation, with different conclusions reached by different authors \citep{pellegrini2012, crocker2012}. Thus, although the systematic $\sim$20-30\% bias is likely to be present for all or most of our regions, we will not attempt to correct for it, in light of the uncertainty just discussed.

\section{Results and Discussions}
\label{results}

\subsection{SFR(H$\alpha$+70)}
\label{hpfir}

\begin{figure*}[h]
\center{
\includegraphics[scale=0.8]{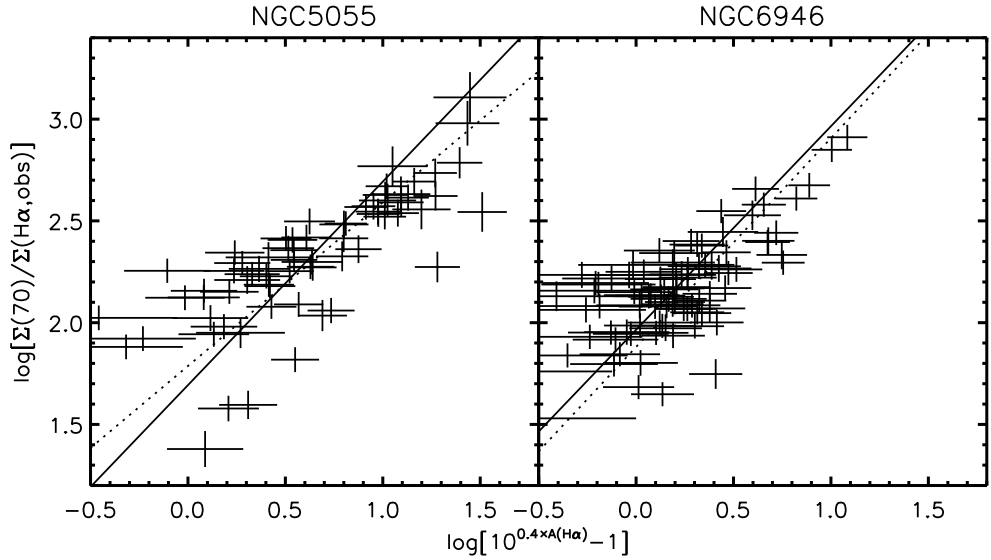}
}
\caption[]{\label{fig:IRvAHa} Ratio of $\rm\Sigma(70)$ to $\rm\Sigma(H\alpha,obs)$ as a function of 10$^{0.4\cdot A(H\alpha)}$-1 (representing the attenuated fraction of H$\alpha$ emission, see section \ref{hpfir}). The two plots are data for NGC\,5055 and NGC\,6946 respectively. Dotted lines are linear fits in log-log space and solid lines are linear fit with unity slope in log-log space.}
\end{figure*}

\citet{calzetti2007} and \citet{kennicutt2007,kennicutt2009} proposed to use a combination of 24~$\mu$m emission and H$\alpha$ emission as an `unbiased' SFR indicator, where the 24~$\mu$m emission is used to represent the attenuated SFR, i.e. attenuated H$\alpha$ emission. Following these studies, we compare the extinction correction and the 70~$\mu$m emission, to see if a similar combination, 
\begin{equation}
\label{equ:fir}
\rm \Sigma(H\alpha,corr)=\Sigma(H\alpha,obs)+\alpha\Sigma(70)
\end{equation}
can be established using the 70~$\mu$m emission instead of the 24~$\mu$m emission, where \textit{corr} stands for extinction corrected. 
Figure \ref{fig:IRvAHa} shows $\rm \Sigma(70)/\Sigma(H\alpha,obs)$ as a function of $\rm10^{0.4A(H\alpha)}-1$, since
\begin{equation}
\rm\alpha=\frac{\Sigma(H\alpha,corr)-\Sigma(H\alpha,obs)}{\Sigma(70)}=\frac{(10^{0.4A(H\alpha)}-1)\Sigma(H\alpha,obs)}{\Sigma(70)}.
\end{equation}

Linear fits in log-log space, $\rm log[\Sigma(70)/\Sigma(H\alpha,obs)]=a_0+b_0\cdot log[10^{0.4A(H\alpha)}-1]$ and linear fits with fixed unity slope in log-log space, $\rm log[\Sigma(70)/\Sigma(H\alpha,obs)]=a_1+log[10^{0.4A(H\alpha)}-1]$, are shown as dotted lines and solid lines respectively and the fit parameters are listed in Table \ref{tab:fit1}. The best-fit slope of NGC\,6946 is consistent with unity and if we adopt the forced fit with unity slope, we obtain $\rm\alpha=1/10^{a_1}$, which is also listed in Table \ref{tab:fit1}. The 1-$\sigma$ dispersions of the data around the linear fit are also listed; these dispersions provide a better measure of the accuracy of the linear fit, as the formal uncertainties listed in Table \ref{tab:fit1} (and subsequent Tables) reflect only the internal dispersion resulting from the fit procedure. We use the IDL routine `fitexy' to perform the fit process here and for later fits. This takes into account the uncertainties in both axes and derives a straight line fit using $\chi$-square minimization. 

We further test the robustness of our fits, as it seems that the linear fit is determined by the highest four data points and thus larger uncertainty may occur without those data points, for the NGC\,6946 fit. Though there is no physical reason to not include these points, we obtain a fit slope of 1.062($\pm$0.111) if we exclude them from the fit, which is not too much different from the original fit slope 1.026($\pm$0.077) (Table \ref{tab:fit1}) and the fit uncertainty is only slightly increased. Thus, it reveals that our fit is actually robust.

\begin{figure*}[h]
\center{
\includegraphics[scale=0.8]{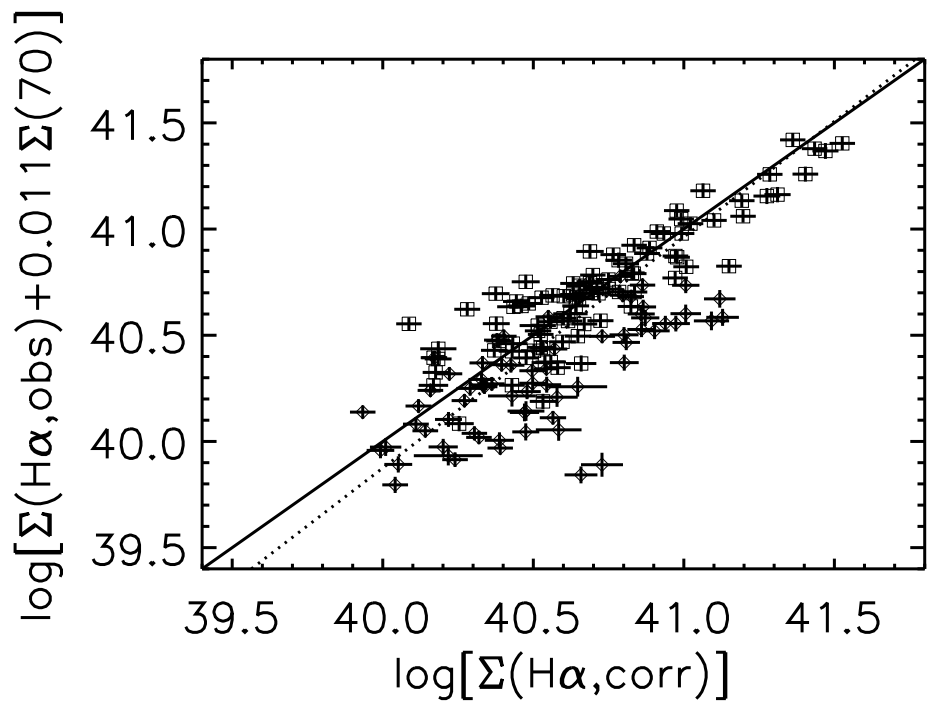}
}
\caption[]{\label{fig:Hap70vB} $\rm\Sigma(H\alpha,obs)+0.011\Sigma(70)$ as a function of extinction corrected $\rm\Sigma(H\alpha)$. The open diamonds and open squares are data for NGC\,5055 and NGC\,6946 respectively. The dotted line is a linear fits in log-log space and the solid line is the 1-to-1 line. The linear fit slope is 1.09(0.02). The dispersion in the data around the linear fit is 0.22 dex. The unit is $\rm ergs\cdot s^{-1}\cdot kpc^{-2}$ for luminosity surface density.}
\end{figure*}

The linear trend obtained for NGC\,6946 indicates that using the PACS 70~$\mu$m band as a complementary SFR tracer to correct for the attenuated H$\alpha$ emission is a viable method, with a corresponding $\alpha$ value of 0.011($\pm$0.001). We obtain a best-fit slope significantly smaller than unity for NGC\,5055, probably due to its larger inclination (see Section \ref{longIR}). To test the applicability of the derived $\alpha$, we plot $\rm log[\Sigma(H\alpha,obs)+0.011\Sigma(70)]$ as a function of $\rm log[\Sigma(H\alpha,corr)]$, the extinction corrected H$\alpha$ luminosity surface density (Fig. \ref{fig:Hap70vB}), for these two galaxies. The general trend of the data is close to the 1-to-1 correlation, indicating that $\alpha=0.011$ works on average. However, the NGC\,5055 data lie systematically below the 1-to-1 correlation. This deviation is discussed in more detail in the next section. A comparison of this derived $\alpha=0.011$ to the proportionality parameter for $\rm H\alpha+\alpha\cdot24~\mu m$ \citep[$\alpha=0.031$,][]{calzetti2007}, gives a ratio L(70)/L(24)$\sim$3, which is consistent with the model predictions by \citet{draine2007a} with an average stellar radiation field strength U about 300, the value estimated for star-forming regions \citep{calzetti2007}. The current sample size is too small to enable a detailed comparison
between the 24 micron and 70 micron mixed indicators, which we defer
to a later paper where a larger number of galaxies will be analyzed.

\subsection{SFR(70)}
\label{sfr70}

In section \ref{hpfir} and in this section we seek to obtain a linear trend between the sum
of H$\alpha$ plus the 70~$\mu$m emission and Br$\gamma$ and between the 70~$\mu$m emission
and Br$\gamma$, respectively. At first sight, it will appear that seeking linear
trends (slope of unity in a log--log plot) in both cases is contradictory. However,
we note that the range of luminosities for which each calibration is
valid will be different and only partially overlapping \citep[see, e.g., discussion in][]{kennicutt2009,calzetti2010}.
 The 70~$\mu$m emission will generally
be stronger in dust obscured regions where H$\alpha$ emission is weak. 

We express the IR (70, 100 and 160~$\mu$m bands) SFR calibrations in this paper as
\begin{equation}
\label{equ:SFR(IR)}
\rm SFR(IR\ band)\ (M_{\odot}\cdot yr^{-1})=\mathcal{C}_{IR\ band,region}\times10^{-43}\times L(IR\ band)\ (ergs\cdot s^{-1})
\end{equation}
We refer to the $\bf \mathcal{C}$ as the Calibration Coefficient. Thus, the Calibration Coefficients in \citet{calzetti2010} and \citet{li2010} are $\rm \mathcal{C}_{70,galaxy}=0.58$, and $\rm \mathcal{C}_{70,700pc}=0.94$ respectively. Using the 70~$\mu$m to the bolometric infrared luminosity ratio, \citet{lawton2010} has derived $\rm \mathcal{C}_{70,10-300pc}=0.97$ for the Magellanic Clouds.

\begin{figure*}[h]
\center{
\includegraphics[scale=0.8]{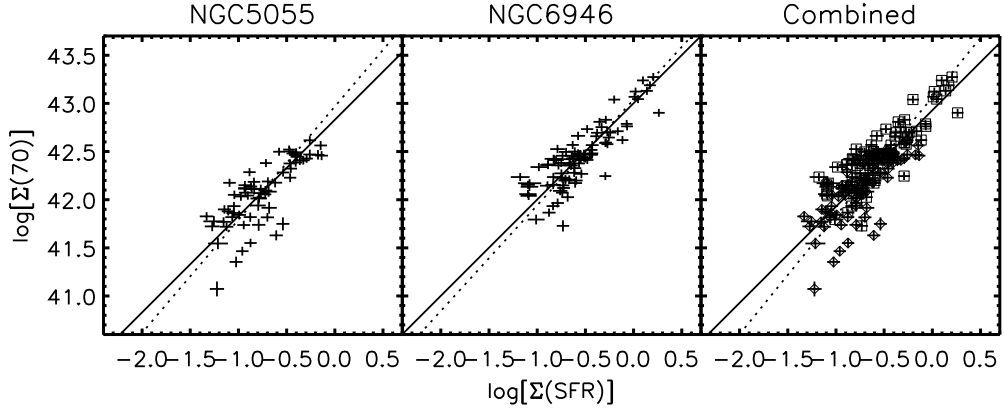}
}
\caption[]{\label{fig:SFR_IR} $\rm\Sigma(70)$ as a function of $\rm\Sigma(SFR)$,  from extinction corrected $\rm\Sigma(Br\gamma)$. The three columns are data for NGC\,5055, NGC\,6946 and both combined respectively. In the third column, open diamonds are for NGC\,5055 and open squares are data for NGC\,6946. Dotted lines are linear fits in log-log space and solid lines are linear fit with unity slope in log-log space. The units are $\rm M_{\odot}\cdot yr^{-1}\cdot kpc^{-2}$ for star formation rate surface density and $\rm ergs\cdot s^{-1}\cdot kpc^{-2}$ for luminosity surface density.}
\end{figure*}

Adopting $\rm SFR=5.45\times10^{-42}\cdot L(H\alpha)$ \citep{calzetti2010} and $\rm L(H\alpha)=103\cdot L(Br\gamma)$, we derive SFRs for our regions using their extinction corrected Br$\gamma$ luminosities. Figure \ref{fig:SFR_IR} shows the $\rm\Sigma(70)$ as a function of $\rm\Sigma(SFR)$. The linear fits in log-log space, $\rm log[\Sigma(70)]=a_0+b_0\cdot log[\Sigma(SFR)]$, and the fixed unity slope fits, $\rm log[\Sigma(70)]=a_1+log[\Sigma(SFR)]$, are shown as dotted lines and solid lines respectively and the fit parameters are listed in Table \ref{tab:fit2}. The 1-$\sigma$ dispersions of the data around the linear fit are also listed.  The fits all have slopes close to unity and so we adopt the fit results with unity slope to obtain the Calibration Coefficients, $\mathcal{C}=10^{43-a_1}$, used in Equation (\ref{equ:SFR(IR)}). These are also listed in Table \ref{tab:fit2}; the  $\mathcal{C}_{70,200pc}=1.18(0.02)$ for the combined data (at $\sim$200 pc scale) will be used for the comparison with the Calibration Coefficients, $\mathcal{C}_{70}$, from \citet{calzetti2010} and \citet{li2010}, as larger samples of galaxies were combined to derive the calibration in those works. 

The calibration coefficient of NGC\,5055 is higher in value than
that of NGC\,6946 (Table \ref{tab:fit1}). We provide analysis with the longer wavelengths and reasons in Section \ref{longIR} that
lead us to believe that the anomalous behavior is in NGC\,5055. We suggest that the higher inclination of NGC\,5055 relative to
NGC\,6946 prevents an effective removal of the diffuse dust along the line
of sight to the H{\sc ii} regions of the galaxy and/or causes over-subtraction of the star forming
 regions contributing to the local 70 micron emission. Although we use circular
annuli around each region to remove the foreground/background diffuse
component of dust emission, the high inclination of NGC\,5055 implies that
there is more intervening dust emission within each photometric
aperture that is not directly related to the H{\sc ii} region(s), and removal of
such a component is subject to larger systematic uncertainties. 
Indeed, as we show in section \ref{longIR}, going to lower resolution (longer wavelengths)
we increase line-of-sight confusion and the NGC\,5055 data further deviate
from those of NGC\,6946. \citet{draine2007b} show that the mean
interstellar radiation field is 2.5 times weaker in NGC\,5055 than in
NGC\,6946, which also manifests itself as a global infrared spectral energy distribution (SED) that has a
slightly colder mean temperature \citep{skibba2011,dale2012}.
Thus, the intervening diffuse component in the NGC\,5055 apertures is made
of dust that is slightly colder than that in NGC\,6946, which accounts for the behavior 
observed in section \ref{longIR} at 100 and 160~$\mu$m. In addition to the possible over-subtraction, 
this also helps account for the higher calibration coefficient in NGC\,5055 than in
NGC\,6946: the peak of the diffuse dust emission in the former galaxy is shifted to longer wavelengths than in the latter. 
In fact, H{\sc ii} regions in NGC\,5055 show higher extinction values than
in NGC\,6946 (Figure \ref{fig:HavB}), which may imply higher self-shielding for the
dust surrounding the H{\sc ii} regions, resulting in cooler mean dust temperatures \citep[e.g.][]{calzetti2010}. Both effects are likely to contribute
to the $\sim$50\% higher $\mathcal{C}_{70,200pc}$ in NGC\,5055. 

We also revisit the data for both NGC\,5055 and NGC\,6946 in \citet{li2010} to understand the $\sim$50\% higher $\mathcal{C}_{70,200pc}$ of NGC\,5055. We find that the NGC\,5055 data in that work already show a 2-$\sigma$ deviation from the mean trend in the sense of being systematically higher than the mean, while the NGC\,6946 data lie about the mean trend. Indeed, the corresponding $\mathcal{C}_{70,700pc}$ value of NGC\,5055 is also about 50\% larger that that of NGC\,6946 in the data of \citet{li2010}. This strongly suggests that a larger sample with more galaxies is needed to average out fluctuations and derive a more reliable general calibration. 

Although we
find roughly linear trends both in section \ref{hpfir} and this section, we should be reminded that
our study only involves two galaxies, and uncertainties are large. A study involving
a larger sample (which we are planning in the near future) will be better suited for
analyzing similarities and differences in the two calibrations.


\subsection{SFR Calibrations at Different Scales}
\label{dscale}

\citet{li2010} found, from a comparison of the Calibration Coefficients between 70~$\mu$m emission in sub--galactic regions ($\rm \mathcal{C}_{70,700pc}$) and for integrated (star-forming and starburst) galaxies ($\rm \mathcal{C}_{70,galaxy}$), a $\sim$40\% difference between the two values, which they attributed to dust heated by stellar populations not related to the current star formation activity. The comparison is performed using the following steps: for the calibration in \citet{calzetti2010}, a galaxy with a SFR of 1 M$_{\odot}\cdot$yr$^{-1}$ implies a 70~$\micron$ luminosity of $1.725\times10^{43}$ ergs$\cdot$s$^{-1}$; the calibration of \citet{li2010} shows that $\sim$700~pc sub--galactic star forming regions with the same total SFR of 1 M$_{\odot}\cdot$yr$^{-1}$ have a total 70~$\micron$ luminosity of only $1.067\times10^{43}$ ergs$\cdot$s$^{-1}$; thus, the difference in these two calibrations reveals an average of $\sim$40\% excess 70~$\micron$ emission in the galaxies on scales larger than 700~pc. With the results in Table \ref{tab:fit2}, we add another $\mathcal{C}_{70}$, derived from the two galaxies, with an average region size of $\sim$210 pc (Fig. \ref{fig:DCal}). $\mathcal{C}_{70}$ increases as we probe star forming regions with smaller size, which gives a factor 2 difference between the 200 pc calibration and the whole galaxy calibration, implying that $\sim$50\% of the 70~$\mu$m emission beyond the 200 pc scale is due to heating of dust by stars not associated with current star formation.

In the comparison above, we assume that the dust heated by young stellar populations in whole galaxies is a scaled-up version of dust heating in 200 pc regions, as traced by the ionizing photon flux. This is equivalent to assuming the UV photons leaked from the H{\sc ii} regions trace the leakage of ionizing photons, and that this ratio is constant over all spatial scales. The leakage of ionizing photons out of H{\sc ii} regions is not included in our estimate of the $\rm\Sigma(SFR)$. If we were to systematically correct all the Br$\gamma$ luminosities by $\sim$30\% (section \ref{other}), the estimate of the diffuse fraction as inferred from the comparison between $\mathcal{C}_{70,200pc}$ and $\mathcal{C}_{70,galaxies}$ would increase by a similar percentage, from $\sim$50\% to $\sim$65\%. This new estimate, however, does not include the leakage of dust heating UV photons from the H{\sc ii} regions; if we assume, as done so far, that they trace the leakage of ionizing photons, the estimate of the diffuse 70~$\mu$m fraction is reset back to $\sim$50\%.

\citet{lawton2010} provide estimates for the size of H{\sc ii} regions in the IR. Their analysis yields  mean 70~$\mu$m radii of 60($\pm$20)~pc for H{\sc ii} regions in the LMC and 80($\pm$30)~pc for H{\sc ii} regions in the SMC. The radii are calculated as encompassing 95\% of the light. Our $\sim$200 pc radii apertures are thus likely to include most of the UV photons responsible for heating the dust within and around H{\sc ii} regions; the 50\% estimate for the diffuse fraction outside 200~pc radii is thus likely to be a conservative one.

The interpretation we give for the trend in Figure \ref{fig:DCal} is that smaller regions contain smaller fractions of diffuse IR emission heated by stellar populations that is not related to current star formation, i.e. populations older than about 10~Myr. We make use of a simple model \citep{li2010} to try to qualitatively support this interpretation. The simple model involves using, a) STARBURST99 to produce a stellar spectral energy distribution for a given star formation history; b) the attenuation curve from \citet{calzetti2000} applied to the stellar light to produce the total IR emission; c) the dust model from \citet{draine2007a} to predict IR emission in a given band (here 70~$\mu$m only). We adopt continuous star formation populations over the three timescales of 100~Myr, 1~Gyr and 10~Gyr and derive $\mathcal{C}_{70}$ for each timescale. The applied attenuation value is the average of the two values from Table \ref{tab:gal}. We then place the three model $\mathcal{C}_{70}$ values on Figure \ref{fig:DCal} (star symbols). The physical scales are assigned to the models based on an assumed region crossing speed of stars of $v\sim$3 km$\cdot$s$^{-1}$  \citep[in agreement with typical velocity dispersions of young stars,][]{sotnikova2003}, a proxy that relates physical scales to timescales. From the crude model comparison, one can tell that the 200~pc calibration (filled circle) is close to the 100~Myr model while the whole galaxy calibration (filled square) is close to a 10~Gyr model. In fact, a longer star formation timescale accumulates more stars with long lifetimes, which will heat the dust. The contribution to the IR emission from these stars will need to be removed from the SFR accounting, resulting in a smaller calibration coefficient for whole galaxies than for small regions.

 At high surface brightness values, the spatially resolved PACS measurements tend
to be higher than the MIPS measurements at the same wavelength, which has been
attributed to the onset of non--linear behavior in the MIPS detectors (see, e.g.,
the photometric experiments by the PACS team\footnote{https://nhscsci.ipac.caltech.edu/pacs/docs/Photometer/PICC-NHSC-TR-034.pdf}$^{,}$\footnote{http://herschel.esac.esa.int/twiki/pub/Public/PacsCalibrationWeb/ExtSrcPhotom.pdf}). If our MIPS 70 $\mu$m measurements
in the 700~pc regions are, indeed, slightly underestimated, we would need to increase
the measured fluxes at this spatial scale, in order to bring them to a consistent
flux scale with the 200~pc measurements. As a result, in order to recover
the correct SFR at 700 pc, the calibration constant would need to be decreased, which
would only increase our already found discrepancy between the 200~pc calibration and the 700~pc calibration. In reality, we expect the effect of the discrepancy between PACS and MIPS responses at high surface brightnesses to be small on 700~pc scales,
since the whole galaxy PACS 70 $\mu$m measurement is not significantly different
from the MIPS one \citep{dale2012}. Thus our estimated diffuse 70 $\mu$m fraction of the total emission, as derived by comparing the 200~pc scale to the integrated calibrations, is likely to be an accurate estimate.

The open circle on Figure \ref{fig:DCal} is the Calibration Coefficient derived from the same combined datasets of NGC\,5055 and NGC\,6946, without the local background subtracted from the 70~$\mu$m photometry. The open diamond is the Calibration Coefficient for $\sim$700~pc regions also without the local background subtracted. Simply by comparing the filled circle and diamond with the open circle and diamond, we obtain about 31\% and 22\% of diffuse emission in the 70~$\mu$m IR bands when comparing whole galaxies with 200~pc and 700~pc regions, respectively. Thus, even when the local background is not subtracted from the 200 pc and 700 pc apertures, we still find significant diffuse emission in the whole galaxy, and its fraction still increases with decreasing aperture size (Fig. \ref{fig:DCal}). The persistency of the trend with aperture size even in the absence of local background subtraction suggests that the diffuse emission contribution in galaxies is significant in the IR bands and supports our choice to remove the local background in our analysis. Our choice of local background subtraction, by using annuli as close as possible to the aperture, is generally more robust than other methods (e.g., large regions backgrounds, etc.) against background variations, both real and spurious. Thus, we consider our results representative of the trends we should recover in other galaxies, as well. However, the local background subtraction is still limited by the resolution and the diffuse contribution associated with scales smaller than the resolution can not be or at least can not be effectively removed, thus producing the differences between the 200 and 700~pc analysis.

To summarize, with the high resolution {\it Herschel} PACS 70~$\mu$m images, we find an even larger $\rm \mathcal{C}_{70}$ for a smaller aperture size ($\sim$200~pc) than previous findings by \citet{li2010} and \citet{calzetti2010} with $\sim$700~pc regions and whole galaxies, respectively, which reveals a diffuse 70~$\mu$m emission fraction of about 50\% beyond our 200~pc regions. We hypothesize this fraction is due to heating of dust by older stars not associated with current star formation. With the comparison to a simple model, we find this increasing trend of $\rm \mathcal{C}_{70}$ with decreasing region sizes to be physically related to the associated star formation timescale, i.e. smaller regions related to a shorter star formation timescale have less fractional contribution from dust heated by stellar populations that are not related to current star formation. 

\subsection{Analysis with 100 and 160~$\mu$m}
\label{longIR}

To perform the analysis at 100 and 160~$\mu$m, we also process the PACS images of these two wavelengths according to the methodology described in section \ref{photometry} (H$\alpha$ and Br$\gamma$ images as well), in order to degrade all images to the same resolution of 6.9$^{\prime\prime}$ for 100~$\mu$m and 11.5$^{\prime\prime}$ for 160~$\mu$m.
The adopted aperture radii are 8$^{\prime\prime}$ for the images at the 100~$\mu$m resolution and 12$^{\prime\prime}$ for the images at the 160~$\mu$m resolution, corresponding to physical sizes of about 330 pc and 400 pc. With these aperture sizes, 65 and 47 regions for NGC\,5055 and 86 and 62 regions for NGC\,6946 are selected at the 100 and 160~$\mu$m resolutions respectively (Fig. \ref{fig:aperfir}). The final median uncertainties (and uncertainty ranges) are about 11\% (10-23\%) at 100~$\mu$m and 12\% (10-26\%) at 160~$\mu$m, while the Br$\gamma$ measurements have similar uncertainties as those mentioned in section \ref{error} at 70~$\mu$m resolution. With the extinction corrected Br$\gamma$ at matched resolution as the reference SFR, the correlations between both 100 and 160~$\mu$m and SFR are shown in Figure \ref{fig:SFR_FIR}. The fit results and derived Calibration Coefficients (where applicable) are listed in Table \ref{tab:fit3}. 

The data for NGC\,6946 show a correlation with slope close to unity between the longer wavelengths of FIR bands and the SFR, and the corresponding $\mathcal{C}_{100}$ and $\mathcal{C}_{160}$ values are derived. However, the best-fit slopes for the NGC\,5055 data deviate significantly from unity, being smaller than 1. Moreover, the slope for the 160~$\mu$m resolution data flattens dramatically. To understand the physical reason behind this flattening, we use the 70~$\mu$m emission as a gauge to study the behavior of longer wavelengths at different IR brightness (Fig. \ref{fig:IRr}). For this analysis, we match the 70~$\mu$m image of the two galaxies to the 100~$\mu$m resolution and to the 160~$\mu$m resolution respectively by performing the same convolution process described above, where each image is convolved with the PSF of the other image. Thus the resolutions of the matched images are about 8.7$^{\prime\prime}$ at 100~$\mu$m and 12.7$^{\prime\prime}$ at 160~$\mu$m, and the adopted aperture radii are 10$^{\prime\prime}$ and 14$^{\prime\prime}$ respectively. The inner and outer annuli radii are again two times and four times the aperture radii. The ratios of $\Sigma(100)/\Sigma(70)$ and $\Sigma(160)/\Sigma(70)$  are plotted as a function of $\Sigma(70)$ for both galaxies, and a linear fit through the data in log-log space is drawn on each panel. For the $\Sigma(160)/\Sigma(70)$ ratio, we use the model mentioned in Section \ref{dscale} (adding the 160~$\mu$m output) to produce a model prediction, shown as dashed lines on the lower panels. The difference of $\Sigma(160)/\Sigma(70)$ ratios between NGC\,5055 and NGC\,6946 is obvious, and the model prediction agrees with the NGC\,6946 data. Since any resolution complication is taken out by matching the resolution among images, the steepening trend of ratios for NGC\,5055 is most likely coming from the fact that this galaxy is more inclined than NGC\,6946 which increases line-of-sight overlap and confusion and/or possible over-subtraction of the star forming regions contributing to the local 70 micron emission. This is further supported by the exacerbation of the effect for increasing wavelength (lower resolution), which increases the line-of-sight confusion. In addition, NGC\,5055 has higher $\Sigma(100)/\Sigma(70)$ and $\Sigma(160)/\Sigma(70)$  ratios on average than NGC\,6946, which agrees with the result that NGC\,5055 has globally colder dust as indicated by \citet{draine2007b} and \citet{dale2012}, and as discussed in section \ref{sfr70}. Thus, the reason for a flat trend of $\Sigma(100)/\Sigma(70)$ and $\Sigma(160)/\Sigma(70)$ ratios versus $\Sigma(70)$ for NGC\,5055, which leads to a flat trend of the 100 and 160~$\mu$m correlation with the SFR, is summarized as: 1) a combination of decreasing resolution and high inclination, which produces higher line-of-sight overlap, and results in a higher degree of uncertainty when removing the diffuse contribution of NGC\,5055 using local background subtraction, and/or causes over-subtraction of the star forming regions contributing to the local 70 micron emission, and/or 2) the regions in NGC\,5055 have colder dust than that in NGC\,6946, possibly a consequence of self-shielding by dust in the H{\sc ii} regions of NGC\,5055, which have high extinction, combined with a globally lower luminosity for the interstellar radiation field (section \ref{sfr70}), and contribute more to the longer wavelengths.

Although we derive $\mathcal{C}_{100}$ and $\mathcal{C}_{160}$ from the NGC\,6946 data, which has reasonably linear behavior, the difference between NGC\,5055 and NGC\,6946 questions the applicability of our Calibration Coefficient to general cases for 100 and 160~$\mu$m. Conservatively, one might assume that the longer wavelengths are not as good SFR indicators as the shorter wavelength 70~$\mu$m emission, but we will revisit this issue with a larger sample of galaxies with Br$\gamma$ imaging.

\section{Summary}
\label{sum}

We use archival CFHT WIRCam Br$\gamma$ narrowband and Ks broadband observations for two galaxies, NGC\,5055 and NGC\,6946, to produce Br$\gamma$ emission line maps. We adopt the Br$\gamma$ emission as our reference SFR indicator for calibrating SFR(70) with the {\it Herschel} PACS 70~$\mu$m data from the KINGFISH observations. The H{\sc ii} region luminosity functions of the two galaxies have a slope consistent with -1, which agrees with previous studies using H$\alpha$, and lends credence to the fact that the Br$\gamma$ map processing, continuum subtraction and absolute photometric calibration are reliable. A comparison with the free-free emission shows the same consistency in the Br$\gamma$ maps.

The SINGS and LVL H$\alpha$ images are used for extinction correction, assuming Case B recombination. Both the Br$\gamma$ and H$\alpha$ images are convolved with the observed {\it Herschel} PSF of the 70~$\mu$m band and the 70~$\mu$m band is also convolved with the Br$\gamma$ PSF, to better match the resolution. Sources are selected based on presence of both IR and Br$\gamma$ emission peaks; thus only the H{\sc ii} regions that are bright enough (to have detected Br$\gamma$ emission) with enough dust (to have IR emission) are selected for our calibrations of SFR(70) and analysis. Aperture photometry is performed on the sources using the standard approach of employing annuli around the photometry apertures to remove the local background.

\begin{figure*}[h]
\center{
\includegraphics[scale=0.8]{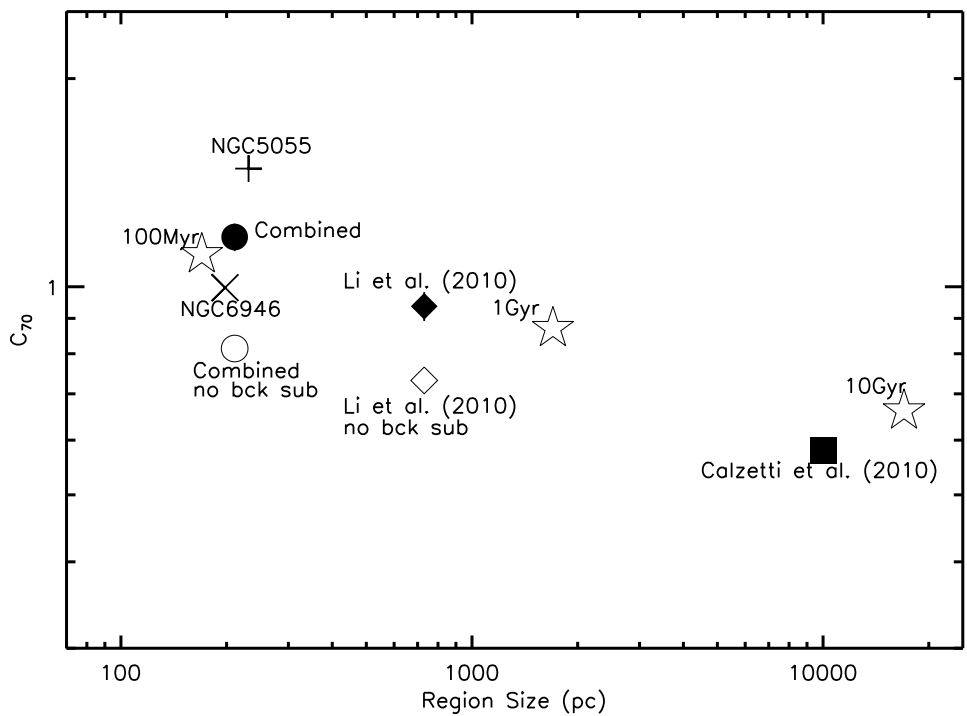}
}
\caption[]{\label{fig:DCal} Calibration Coefficients of the 70~$\mu$m band ($\mathcal{C}_{70}$) as a function of region physical size. The plus symbol is for NGC\,5055 and the cross symbol is for NGC\,6946, using the values derived in the present work. The filled circle is for both galaxies combined (with a combined sample of 160 regions). The filled diamond is from \citet[using 556 regions]{li2010}. The filled square is from \citet[using 189 star-forming and starburst galaxies]{calzetti2010}. For all data with filled symbols, 3-$\sigma$ error bars are shown (the filled square is larger than its error bar). The open circle is for the two galaxies in our work combined, but without the local background removed from the 70~$\mu$m photometry. Similarly the open diamond is for the 700~pc regions of \citet{li2010} without the local background removed. The three star symbols are $\mathcal{C}_{70}$ determined for continuous star formation populations with different timescales, using a simple model; they are placed on the plot after associating physical scales to their duration of star formation (see text).  }
\end{figure*}

\begin{figure*}[h]
\center{
\includegraphics[scale=0.8]{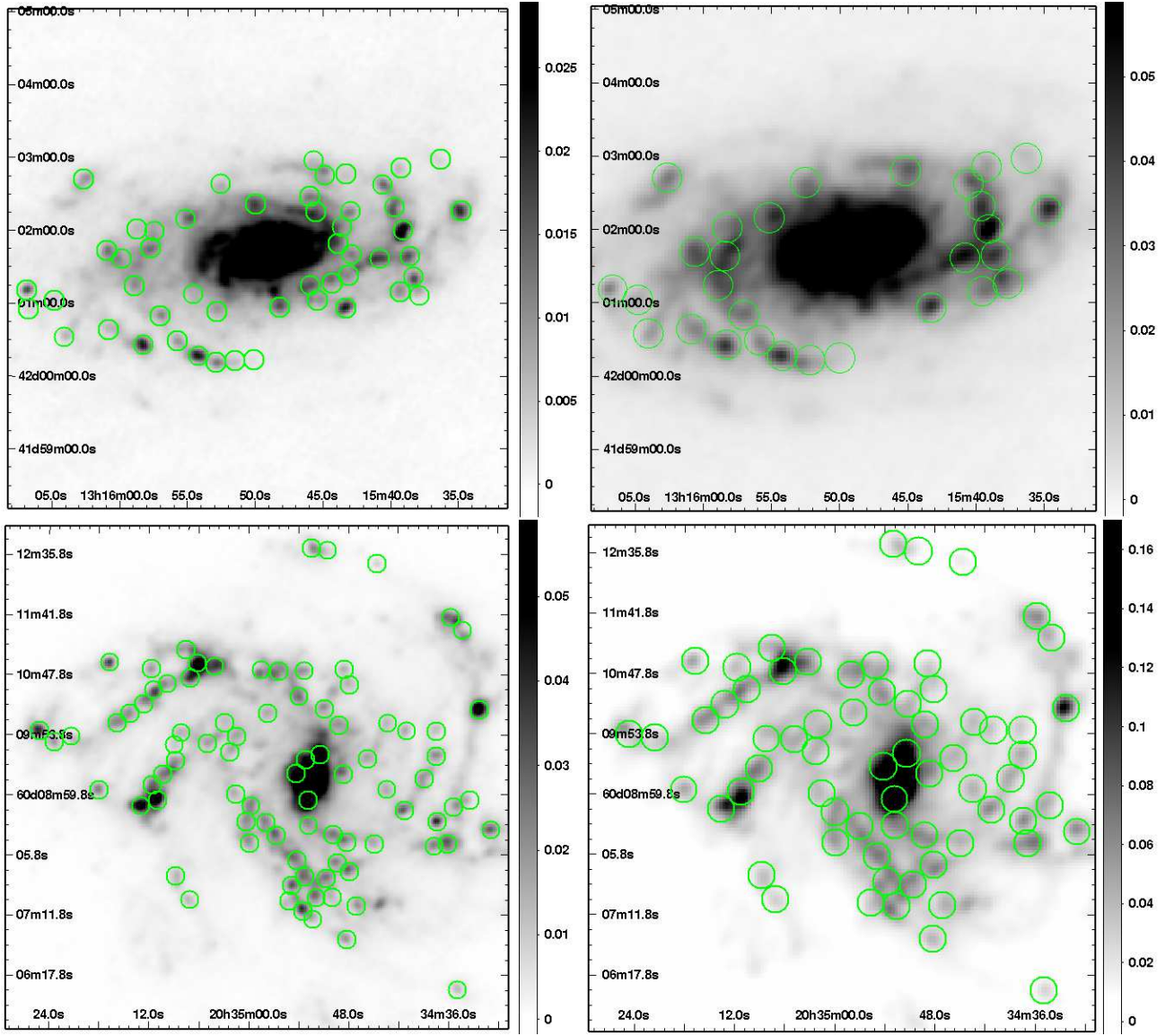}
}
\caption[]{\label{fig:aperfir} From top left to bottom right: 100~$\mu$m and 160~$\mu$m images for NGC\,5055 and 100~$\mu$m and 160~$\mu$m images for NGC\,6946. The green circles are the apertures we adopted for 100~$\mu$m and 160~$\mu$m resolution analysis, respectively. North is up and east is left. The image size is about 410$^{\prime\prime}\times$410$^{\prime\prime}$ (about 15.8$\times$15.8 kpc) for each panel of NGC\,5055 and about 450$^{\prime\prime}\times$450$^{\prime\prime}$ (about 14.8$\times$14.8 kpc) for each panel of NGC\,6946. The convolved Br$\gamma$ images are not shown but the uneven background is worse with lower resolution. The units for the grayscale bars are all Jy per pixel, while the 100~$\mu$m images have pixel sizes of 1.7$^{\prime\prime}$ and the 160~$\mu$m images have pixel sizes of 2.85$^{\prime\prime}$.}
\end{figure*}

\begin{figure*}[h]
\center{
\includegraphics[scale=0.8]{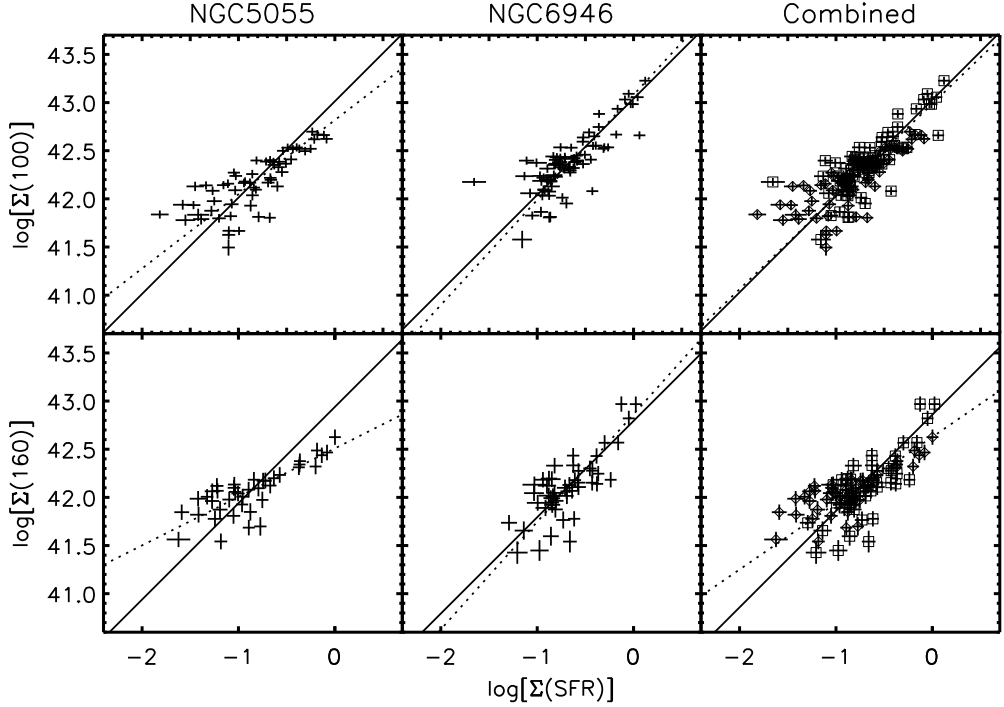}
}
\caption[]{\label{fig:SFR_FIR} $\rm\Sigma(100)$ and $\rm\Sigma(160)$ as a function of $\rm\Sigma(SFR)$,  from extinction corrected $\rm\Sigma(Br\gamma)$. The vertical panels are data for NGC\,5055 (left), NGC\,6946 (middle) and both combined (right), respectively. In the right panel, open diamonds are for NGC\,5055 and open squares are for NGC\,6946. Dotted lines are linear fits in log-log space and solid lines are linear fit with unity slope in log-log space. We notice the flattening trend of NGC\,5055 with longer wavelength, which is mainly due to a combined effect of decreasing resolution and higher inclination for this galaxy than NGC\,6946. The units are $\rm M_{\odot}\cdot yr^{-1}\cdot kpc^{-2}$ for star formation rate surface density and $\rm ergs\cdot s^{-1}\cdot kpc^{-2}$ for luminosity surface density separately.}
\end{figure*}

\begin{figure*}[h]
\center{
\includegraphics[scale=1]{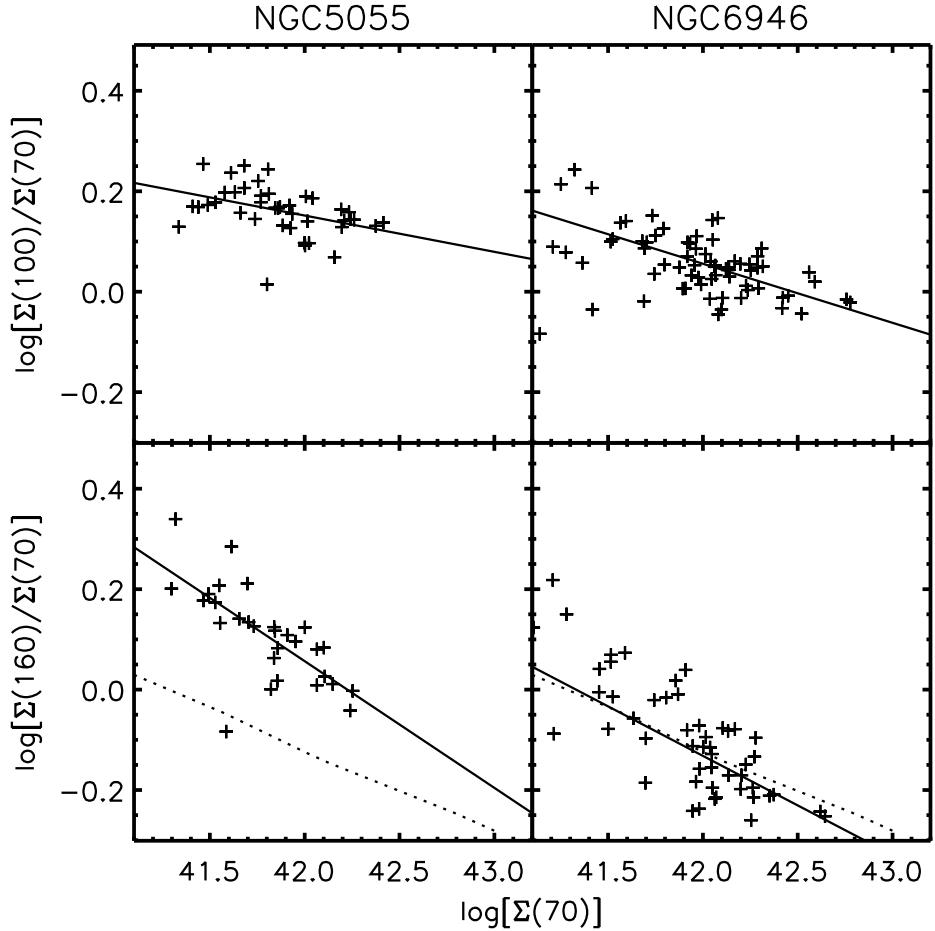}
}
\caption[]{\label{fig:IRr} The ratios of $\rm\Sigma(100)$ and $\rm\Sigma(160)$ to $\rm\Sigma(70)$ as a function of $\rm\Sigma(70)$. The solid lines are linear fits through the data in log-log space. Error bars are shown for each data point. The ratios for NGC\,5055 are on average higher than the ratios for NGC\,6946. The dashed lines on the lower panels are the same expectation from the \citet{draine2007a} model, which agrees with the NGC\,6946 data. The NGC\,5055 ratio of 160~$\mu$m to 70~$\mu$m is much steeper than the model prediction, and is most likely related to the higher inclination of NGC\,5055, since the resolutions are matched between 70~$\mu$m and 160~$\mu$m for this analysis. The corresponding U, the average stellar radiation field strength defined by \citet{draine2007a} and adopted in the model, are about 0.94 to 3.07 and 1.50 to 4.97 for the range of $\rm\Sigma(160)/\Sigma(70)$ ratio for NGC\,5055 and NGC\,6946 respectively. The unit is $\rm ergs\cdot s^{-1}\cdot kpc^{-2}$ for luminosity surface density.}
\end{figure*}

We use all sources with S/N$>$3 at a depth of L(Br$\gamma$)$\sim$10$^{36.6}$ which corresponds to star clusters with mass $\sim$ 3-6$\times$10$^{4}$ M$_{\odot}$, assuming an average age of 4-5 Myr and a \citet{kroupa2001} stellar IMF. With the estimates of \citet{cervino2002} adapted to our IMF, it gives a fluctuation in the ionizing photon flux less than 30\%. Thus random fluctuation in the ionizing photon flux is not a concern. Our analysis finds:
\begin{enumerate}
\item The 70~$\mu$m emission can be combined with the H$\alpha$ emission to produce an unbiased SFR indicator that combines dust obscured and unobscured star formation. The two luminosity surface densities are combined as $\Sigma_{H\alpha,obs}+(0.011\pm0.001)\Sigma(70)$, where the proportionality constant between the observed H$\alpha$ and 70~$\mu$m emission is derived for NGC\,6946 (Table \ref{tab:fit1}). The same constant for NGC\,5055 would be higher, but this galaxy suffers from strong inclination effects (Section \ref{results}).
\item Correlations between the $\rm\Sigma(70)$ and $\rm\Sigma(SFR)$ give the Calibration Coefficient estimate, $\mathcal{C}_{70,200pc}=1.18(\pm0.02)$ (Table \ref{tab:fit2}), after combining data from both galaxies. Comparison between the derived $\mathcal{C}_{70}$ in this work and those from \citet{calzetti2010} and \citet{li2010} reveals a trend of increasing $\mathcal{C}_{70}$ with decreasing region size (Fig. \ref{fig:DCal}), implying that in galaxies about 50\% of the 70~$\mu$m emission outside of a 200 pc scale centered on bright H{\sc ii} regions is unrelated to current star formation. The trend is still present even when the local background is not removed from the 70~$\mu$m photometry. Further analysis with a simple model comparison links this trend to a relation with different star formation timescales, i.e. larger regions have longer star formation timescales and thus have more IR emission produced by dust heated by stellar populations that are not related to current star formation as traced by the ionizing photons. This result, first discussed in \citet{li2010}, is confirmed here by exploiting a larger range of physical sizes.
\end{enumerate}

Whenever applying a SFR(IR) calibration, it is essential to choose a suitable SFR(IR) at the relevant physical scale (star formation timescale) of the system. As shown by \citet{li2010} (Equation (5) in the paper), a metallicity change from about solar to extreme sub-solar could mean a factor 2 change in the Calibration Coefficient, which should also be taken into consideration when dealing with low metallicity systems.

A similar analysis for PACS 100 and 160~$\mu$m is presented in Section \ref{longIR}. Although the results are limited due to the smaller number statistics with lower resolution and the dispersions in the data are similar to or slightly larger than that of the 70~$\mu$m, the increased variations between the two galaxies at longer wavelengths hint that they are possibly not as good SFR indicators as 70~$\mu$m due to more significant contributions from dust heated by stellar populations that are not related to current star formation and its variations in different galaxies. This is in agreement with the conclusions of \citet{calzetti2010} for whole galaxies and \citet{lawton2010} for H{\sc ii} regions in the LMC and SMC. Ultimately, given the differences we find in the calibration constants of the two galaxies analyzed in this study, we will need to investigate a larger sample, in order to confirm the mean values derived in this work. We expect to perform the same analysis using a 10-fold larger sample in the near future. 

\acknowledgments

This work is based in part on observations made with  {\it Herschel}, a European
Space Agency Cornerstone Mission with significant participation by NASA.
Partial support for this work was provided by NASA through the award 1369560
issued by JPL/Caltech. This work has also been partially supported by the
NASA--ADAP grant NNX10AD08G.
The research of C.D.W. is supported by grants from the Natural Sciences and Engineering Research Council of Canada.

The authors would like to thank Daniel Devost for the WIRCam observations with RUNID 07AD86 for NGC\,5055 and 07BD91 for NGC\,6946.

This work has made use of the NASA/IPAC Extragalactic Database (NED), which is operated by the Jet Propulsion Laboratory, California Institute of Technology, under contract with the National Aeronautics and Space Administration. 

The authors thank an anonymous referee for valuable comments that have helped improve this paper.

\begin{deluxetable}{cccc}
\tablecolumns{4}
\tablewidth{0pc}
\tablecaption{\label{tab:data} Multi-wavelength Data}
\tablehead{
\colhead{Data} & \colhead{Instrument} & \colhead{Central Wavelength} & PSF\\
(1) & (2) & (3) & (4)
}
\startdata
Br$\gamma$ & CFHT WIRCam & 2.166~$\mu$m & $\sim$1.1$^{\prime\prime}$ \\
Ks & CFHT WIRCam & 2.146~$\mu$m & $\sim$1$^{\prime\prime}$ \\
H$\alpha$ & KPNO 2.1m/2.3m Bok & 6573\AA & $\sim$1$^{\prime\prime}$ \\
70~$\mu$m & {\it Herschel} PACS & 70~$\mu$m & 5.5$^{\prime\prime}$ \\
100~$\mu$m & {\it Herschel} PACS & 100~$\mu$m & 6.9$^{\prime\prime}$ \\
160~$\mu$m & {\it Herschel} PACS & 160~$\mu$m & 11.5$^{\prime\prime}$ \\
Free-free emission & GBT Ka-band & 33 GHz & 25$^{\prime\prime}$ \\
\enddata
\tablecomments{col 1. data name; col 2. telescope, instrument and filter; col 3. filter's central wavelength; col 4. FWHM of the PSF.}
\end{deluxetable}

\begin{deluxetable}{cccccccccc}
\tablecolumns{10}
\tablewidth{0pc}
\tabletypesize{\small}
\tablecaption{\label{tab:gal} Sample}
\tablehead{
\colhead{Galaxy} & \colhead{Type} & \colhead{Distance} & \colhead{Diameter} & \colhead{\textit{i}} & \multicolumn{2}{c}{12+log(O/H)} & \colhead{[NII]/H$\alpha$}  & \multicolumn{2}{c}{E(B-V)} \\
& & (Mpc) & ($^{\prime}\times^{\prime}$) & (deg) & (PT) & (KK) & & Foreground & Regions \\
(1) & (2) & (3) & (4) & (5) & (6) & (7) & (8) & (9) & (10)
}
\startdata
NGC\,5055 & SAbc & 7.94 & 12.6$\times$7.2 & 56 & 8.40 & 9.14 & 0.486$\pm$0.019 & 0.018 & 0.65  \\
NGC\,6946 & SABcd & 6.8 & 11.5$\times$9.8 & 32 & 8.40 & 9.05 & 0.448$\pm$0.087 & 0.303 & 0.44  \\
\enddata
\tablecomments{col 1. galaxy name; col 2. galaxy type; col 3. galaxy distance from \citet{kennicutt2011}; col 4. galaxy sizes from \citet{kennicutt2011}; col 5. inclination from \citet{moustakas2010}; col 6-7. characteristic oxygen abundances of the galaxies; the two columns, (PT) and (KK), are the two oxygen abundances listed in Table 9 of \citet{moustakas2010}: the
PT value, in the left-hand-side column, is from the empirical calibration of \citet{pt05}; the KK value, in the right-hand-side column, is from the theoretical calibration of \citet{kk04}; col 8. [NII]/H$\alpha$ ratio from \citet{kennicutt2009}; col 9. E(B-V) used for foreground Galactic extinction correction from NED \citep{schlegel1998,schlafly2011}; col 10. median E(B-V) derived from H$\alpha$ and Br$\gamma$ luminosity for selected H{\sc ii} regions at 70~$\mu$m resolution, after removal of the Galactic foreground extinction \citep{schlegel1998, odonnell1994} . }
\end{deluxetable}

\begin{deluxetable}{cccccc}
\tablecolumns{6}
\tablewidth{0pc}
\tablecaption{\label{tab:file} Archival CFHT Images}
\tablehead{
& \multicolumn{2}{c}{Ks} & \multicolumn{2}{c}{Br$\gamma$} & Note\\
\colhead{Galaxy} & \colhead{\#} & \colhead{Exposure Time (s)} & \colhead{\#} & \colhead{Exposure Time (s)}  & \\
(1) & (2) & (3) & (4) & (5) & (6)
}
\startdata
NGC\,5055 & 112 & 20 & 99* & 100 & RUNID:07AD86 \\
NGC\,6946 & 61 & 10 & 61 & 110 & RUNID:07BD91 \\
\enddata
\tablecomments{col 1. galaxy name; col 2-3. number of images and exposure time with the Ks broadband filter; col 4-5. number of images and exposure time with the Br$\gamma$ narrowband filter; col 6. observation information. *12 of the narrow--band exposures for NGC5055 have exposure times
of 200 seconds, and we rescale them to 100--seconds exposure maps. }
\end{deluxetable}

\begin{deluxetable}{cccccc}
\tablecolumns{6}
\tablewidth{0pc}
\tablecaption{\label{tab:fit1} Fitting results: $\rm log[\Sigma(70)/\Sigma(H\alpha,obs)]$ vs. $\rm log[10^{0.4\cdot A(H\alpha)}-1]$ }
\tablehead{
\colhead{} & \multicolumn{3}{c}{Fitting Parameters} \\
\cline{2-4}
\colhead{} & \colhead{a$_0$} & \colhead{b$_0$} & \colhead{a$_1$} & \colhead{$\alpha$} & \colhead{$\sigma$} \\
(1) & (2) & (3) & (4) & (5) & (6) 
}
\startdata
NGC\,5055 & 1.787(0.034) & 0.804(0.049) & 1.693(0.021) & ... & 0.23 \\
NGC\,6946 & 1.881(0.027) & 1.026(0.077) & 1.965(0.024) & 0.011(0.001) & 0.24 \\
\enddata
\tablecomments{col.1 galaxy; col 2-4. fit parameters as in y=a$_0$+b$_0$x and y=a$_1$+x; col 5. derived $\alpha$ assuming $\rm \Sigma_{H\alpha,corr}=\Sigma_{H\alpha,obs}+\alpha\Sigma_{70}$ with $\alpha$=1/10$\rm^{a_1}$ (see section \ref{hpfir}); col 6. 1-$\sigma$ dispersion of the data around the linear fit. }
\end{deluxetable}

\begin{deluxetable}{cccccc}
\tablecolumns{6}
\tablewidth{0pc}
\tablecaption{\label{tab:fit2} Fitting results: log$\rm[\Sigma(70)]$ vs. log[$\rm\Sigma(SFR)$]}
\tablehead{
\colhead{} & \multicolumn{3}{c}{Fitting Parameters} \\
\cline{2-4}
\colhead{} & \colhead{a$_0$} & \colhead{b$_0$} & \colhead{a$_1$} & \colhead{$\mathcal{C}_{70}$} & \colhead{$\sigma$} \\
(1) & (2) & (3) & (4) & (5) & (6) 
}
\startdata

NGC\,5055 & 42.967(0.037) & 1.176(0.047) & 42.829(0.010) & 1.48(0.04) & 0.24 \\
NGC\,6946 & 43.038(0.020) & 1.097(0.033) & 43.001(0.009) & 1.00(0.02) & 0.19 \\
Combined & 43.065(0.019) & 1.232(0.027) & 42.928(0.007) & 1.18(0.02) & 0.23 \\

\enddata
\tablecomments{col.1 dataset; col 2-4. fit parameters as in y=a$_0$+b$_0$x and y=a$_1$+x; col 5. derived Calibration Coefficient, $\mathcal{C}_{70}=10^{43-a_1}$; col 6. 1-$\sigma$ dispersion of the data around the linear fit.}
\end{deluxetable}

\begin{deluxetable}{cccccc}
\tablecolumns{6}
\tablewidth{0pc}
\tablecaption{\label{tab:fit3} Fitting results: log$\rm[\Sigma(100)]$ \& log$\rm[\Sigma(160)]$ vs. log[$\rm\Sigma(SFR)$]}
\tablehead{
\colhead{} & \multicolumn{3}{c}{Fitting Parameters} \\
\cline{2-4}
\colhead{} & \colhead{a$_0$} & \colhead{b$_0$} & \colhead{a$_1$} & \colhead{$\mathcal{C}$} & \colhead{$\sigma$} \\
(1) & (2) & (3) & (4) & (5) & (6) 
}
\startdata
\cutinhead{100~$\mu$m}
NGC\,5055 & 42.818(0.026) & 0.771(0.028) & 43.013(0.011) & ... & 0.20 \\
NGC\,6946 & 43.079(0.027) & 1.091(0.037) & 43.041(0.010) & 0.91(0.02) & 0.21 \\
Combined & 42.987(0.018) & 0.963(0.023) & 43.029(0.007) & 0.94(0.02) & 0.22 \\

\cutinhead{160~$\mu$m}
NGC\,5055 & 42.508(0.034) & 0.503(0.035) & 42.938(0.017) & ... & 0.15 \\
NGC\,6946 & 42.863(0.050) & 1.117(0.068) & 42.793(0.017) & 1.61(0.06) & 0.22 \\
Combined & 42.628(0.026) & 0.688(0.030) & 42.859(0.012) & ... & 0.20 \\

\enddata
\tablecomments{col.1 dataset; col 2-4. fit parameters as in y=a$_0$+b$_0$x and y=a$_1$+x; col 5. derived Calibration Coefficient, $\mathcal{C}=10^{43-a_1}$; col 6. 1-$\sigma$ dispersion of the data around the linear fit.}
\end{deluxetable}

\end{document}